\documentclass[hyper,a4paper]{JHEP3}
   
\usepackage{epsfig}
\usepackage{latexsym}
\usepackage{graphicx}
\usepackage{amsmath}
\usepackage{amsfonts}   
\usepackage{amssymb}    
\usepackage{float}
\usepackage{bm}
\usepackage{cite}
\newcommand{\comment}[1]{}
\newcommand{\newc}{\newcommand}

\def\thetab0{\theta_{B_0}}

\def\r2{\sqrt 2}
\def\beq{\begin{equation}}
\def\eeq{\end{equation}}
\def\bea{\begin{eqnarray}}
\def\eea{\end{eqnarray}}
\def\ovl{\overline}
\def\sinW2{\sin^2\theta_W}

\def\mz2{M_{z}^2}
\def\c2b{\cos 2\beta}

\def\mz{M_Z}

\def\bN{\mathbf N}
\def\bU{\mathbf U}
\def\bV{\mathbf V}
\def\bRs{\mathbf R^{S^0}}
\def\bRp{\mathbf R^{P^0}}
\def\bRc{\mathbf R^{S^{\pm}}}

\def\lm{\lambda}
\def\kp{\kappa}

\def\n{{\widetilde \chi}^0}

\def\q{\widetilde q}
\def\u{\widetilde u}
\def\d{\widetilde d}
\def\Rsu{\mathbf R^{\widetilde u}}
\def\Rsd{\mathbf R^{\widetilde d}}
\def\Ru{\mathbf R^{u}}
\def\Rd{\mathbf R^{d}}
\def\ps{\displaystyle{\not} p}

\def\sec2w{sec^2\theta_W}

\def\gmin2{(g-2)_\mu}

\catcode`\@=11 

\def\lsim{\mathrel{\mathpalette\@versim<}}
\def\gsim{\mathrel{\mathpalette\@versim>}}
\def\@versim#1#2{\vcenter{\offinterlineskip
    \ialign{$\m@th#1\hfil##\hfil$\crcr#2\crcr\sim\crcr } }}

\newc{\ra}{\rightarrow}
\newc{\s}{\smallskip}
\newc{\nn}{\noindent}
\newc{\non}{\nonumber}

\def \chonep{{\wt\chi_1}^{+}}
\def \chonem{{\wt\chi_1^-}}
\def \chonep2{{\wt\chi_2^+}}
\def \chonem2{{\wt\chi_2^-}}

\title
{Radiative contribution to neutrino masses and mixing in $\mu \nu$SSM}
\author{Pradipta Ghosh,$^a$ ~Paramita Dey,$^{b}$ ~Biswarup Mukhopadhyaya$^{a,c}$ 
and Sourov Roy$^a$\\
$^a$Department of Theoretical Physics and Centre for Theoretical
Sciences,\\
 Indian Association for the Cultivation of Science, \\
2A $\&$ 2B Raja S.C. Mullick Road, Kolkata 700 032, India\\

$^b$Institut f$\ddot{u}$r
Theoretische Teilchenphysik und Kosmologie, \\
RWTH Aachen University, \\
D-52056 Aachen, Germany\\

$^c$Regional Centre for Accelerator-based Particle Physics, \\
Harish-Chandra Research Institute, \\
Chhatnag Road, Jhusi, Allahabad--211019, India\\

E-mails:\email{ tppg@iacs.res.in, paramita@physik.rwth-aachen.de, biswarup@mri.ernet.in,  tpsr@iacs.res.in
}}

\abstract{In an extension of the minimal supersymmetric standard model
  (popularly known as the $\mu \nu$SSM), three right handed neutrino
  superfields are introduced to solve the $\mu$-problem and to
  accommodate the non-vanishing neutrino masses and mixing. Neutrino
  masses at the tree level are generated through $R-$parity violation
  and seesaw mechanism. We have analyzed the full effect of one-loop 
  contributions to the neutrino mass matrix. We show that the 
  current three flavour global neutrino data can be accommodated in 
  the $\mu\nu$SSM, for both the tree level and one-loop corrected analyses.
  We find that it is relatively easier to accommodate the normal 
  hierarchical mass pattern compared to the inverted hierarchical or 
  quasi-degenerate case, when one-loop corrections are included.}

\preprint{TTK-10-17\\
HRI-RECAPP-2010-003}

\begin{document}
\section{Introduction}
\label{Introduction}
Despite its stupendous success in explaining elementary particle interactions,
the celebrated standard model (SM) of particle physics suffers from some
shortcomings, both theoretical and experimental. On the experimental side,
explaining the masses and the mixing pattern of neutrinos is a task in which
the SM is an apparent failure. The neutrino sector, therefore, is a natural
testing ground for most proposals for going beyond the SM. 

Supersymmetry (SUSY) is a rather popular choice for new physics. The minimal 
supersymmetric version of the SM (MSSM) provides a natural solution to the 
so-called ``gauge-hierarchy problem'' through the introduction of superpartners 
of SM-particles. However, MSSM itself is not free of drawbacks. One of these 
is the so-called $\mu$-problem \cite{Kim-Niles}, which essentially means our lack 
of understanding as to why the higgsino mass parameter, a SUSY invariant quantity,
has to be around the SUSY breaking scale. This problem can be solved 
in the next-to minimal supersymmetric version of the standard model (NMSSM), where, 
unlike in the MSSM, the $\mu$-term becomes a derived quantity with the right order 
of magnitude. 

Unfortunately, neither the MSSM nor NMSSM by itself can explain the observed 
pattern of neutrino masses and mixing, on which definite guidelines have been set 
down by existing data \cite{boris_review_08, Schwetz-Valle}. The situation 
becomes different if one allows violation of the discrete symmetry known as 
$R$-parity ($R_p$) 
\cite{r-parity-mssm,Aulakh-Mohapatra,Hall-Suzuki,Lee,Ross-Valle,Ellis-ross-others,masiero-valle,Allanach-Dedes-Dreiner,
r-parity-review}, 
defined as $R_p = (-1)^{L + 3B + 2S}$, where 
$L(B)$ is the lepton(baryon) number and $S$ is the
spin, of a particle. Neutrino masses and mixing have been analyzed in these models under
various assumptions, both at the tree level and by taking loop-induced effects into 
account. Neutrino mass generation in variants of $R_p$ violating MSSM have
been addressed in 
refs. \cite{dawson,dimopoulos-hall,godbole-roy-tata,drees-pakvasa-tata-veldhuis,rakshit-bhattacharyya-raychaudhuri,
adhikari-omanovic,Borzumati-Lee,paramita-biswarup-anirban-soumitra,Joshipura-Nowakowski,Nowakowski-Pilaftsis,
Borzumati-Grossman-others,Banks-Grossman,Hempfling-Rpv,deCarlos-White,Nardi,nilles-polonsky,de-campos-joshipura,
Roy-Mukhopadhyaya,diaz-romao-valle,Roy-Mukhopadhyaya-Datta,bisset-kong-macesanu-orr,Choi-Chun-Hwang,Choi-kang-others,
Joshipura-Vempati,Kaplan-Nelson,Takayama-Yamaguchi,Grossman-Haber-prd,Grossman-Haber,chun-kang,Hirsch-loop-FD,
Valle-Porod-others,Davidson-Losada-jhep,Davidson-Losada-prd,Abada-Davidson-Losada,Abada-Bhattacharyya-Losada,Davidson-Losada-Rius,
Chun-Jung-Park,Grossman-Rakshit-prd,Dedes-Rimmer-Rosiek,Roy-Mukhopadhyaya-Vissani,choi-chun-kang-lee,romao-diaz-hirsch-porod-valle,
datta-mukhopadhyaya-vissani,porod-hirsch-romao-valle,chun-jung-kang-park,jung-kang-park-chun,romao-santos-valle,
giudice-masiero-pietroni-riotto,umemura-yamamoto,Choubey-Mitra}. 
Such extensive study has established $R_p$-violation to be as potent in neutrino mass generation as the well-known seesaw mechanism
\cite{Minkowski:1977sc,GellMann-Ramond-Slansky,Yanagida:1980,Glashow:1979vf,Mohapatra-Senjanovic} 
which requires introduction of gauge singlet neutrino superfields.

R-parity violation, in both the contexts of accelerator phenomenology and, for example, neutrino mass 
generation, have been extensively studied in various scenarios, especially in its L-violating incarnation.
Thus one has so-called trilinear R-parity violation driven by the $\lambda_{ijk}$ or 
$\lambda^\prime_{ijk}$-type terms in the superpotential. In addition to new signals induced by three-body
decays of the lightest neutralino, neutrino masses are generated through loop effects in such 
scenarios \cite{dawson,dimopoulos-hall,godbole-roy-tata,drees-pakvasa-tata-veldhuis,rakshit-bhattacharyya-raychaudhuri,
adhikari-omanovic,Borzumati-Lee,paramita-biswarup-anirban-soumitra}. 
Then one can have R-parity broken by bilinear terms \cite{Joshipura-Nowakowski,Nowakowski-Pilaftsis,
Borzumati-Grossman-others,Banks-Grossman,Hempfling-Rpv,deCarlos-White,Nardi,nilles-polonsky,de-campos-joshipura,
Roy-Mukhopadhyaya,diaz-romao-valle,Roy-Mukhopadhyaya-Datta,bisset-kong-macesanu-orr,Choi-Chun-Hwang,Choi-kang-others,
Joshipura-Vempati,Kaplan-Nelson,Takayama-Yamaguchi,Grossman-Haber-prd,Grossman-Haber,chun-kang,Hirsch-loop-FD,
Valle-Porod-others,Davidson-Losada-jhep,Davidson-Losada-prd,Abada-Davidson-Losada,Abada-Bhattacharyya-Losada,Davidson-Losada-Rius,
Chun-Jung-Park,Grossman-Rakshit-prd,Dedes-Rimmer-Rosiek,Roy-Mukhopadhyaya-Vissani,choi-chun-kang-lee,romao-diaz-hirsch-porod-valle,
datta-mukhopadhyaya-vissani,porod-hirsch-romao-valle,chun-jung-kang-park,jung-kang-park-chun} of the type $\epsilon_i L_i H_2$. 
One here notices remarkable features like (a) non-zero vacuum expectation values for sneutrinos, and (b) the mixing between 
neutrinos and neutralinos as well as charged leptons and charginos. Such a scenario generates one neutrino mass at the tree level 
while the other mass(es) need to be generated via loop effects \cite{Grossman-Haber-prd,Grossman-Haber,chun-kang,Hirsch-loop-FD,
Valle-Porod-others,Davidson-Losada-jhep,Davidson-Losada-prd,Abada-Davidson-Losada,Abada-Bhattacharyya-Losada,Davidson-Losada-Rius,
Chun-Jung-Park,Grossman-Rakshit-prd,Dedes-Rimmer-Rosiek}. The characteristic signal consists in final state with comparable 
numbers of muons and taus at high energy colliders \cite{Roy-Mukhopadhyaya-Vissani,choi-chun-kang-lee,romao-diaz-hirsch-porod-valle,
datta-mukhopadhyaya-vissani,porod-hirsch-romao-valle,chun-jung-kang-park,jung-kang-park-chun}.  

Bilinear R-parity violation, is also linked with spontaneous L-breaking \cite{masiero-valle,romao-santos-valle,
giudice-masiero-pietroni-riotto,umemura-yamamoto,Choubey-Mitra} via a singlet sneutrino
vacuum expectation value (VEV). Such an effect triggers terms of the type $\epsilon_i L_i H_2$ in the
superpotential, and also leaves as its footprint a Majoron which has its own experimental signature 
\cite{gonzalez-garcia-romao-valle,adhikari-mukhopadhyaya,hirsch-vicente-porod}, being an additional source of missing energy. 

In the backdrop of such a rich phenomenology of relatively minimalistic R-parity breaking models, further 
embellishments on the minimal scenarios have also been studied, often with some specific goals. 
A proposal for neutrino mass generation together with a solution to the $\mu$-problem, with the same set 
of gauge-singlet right chiral neutrino superfields, has been advocated in \cite{munoz-lopez_fogliani}. This 
is popularly known as $\mu \nu$SSM. Following this proposal, the scalar sector and the parameter space of 
$\mu \nu$SSM was studied in great detail in \cite{munoz-lopez-2}. In $\mu \nu$SSM, the three generations of 
SM-neutrinos can acquire masses through a TeV-scale seesaw mechanism, with both neutralinos and heavy neutrinos 
participating in the process \cite{Ghosh-Roy}. Issues of neutrino mass generation and the $\mu$-problem can 
also be found in some recent works \cite{kitano-oda,frank-huitu-ruppell,chemtob-pandita,gautam-moreau-abada,Hundi-Tata-Pakvasa}.

A comprehensive analytical study of mass generation of light neutrinos
in $\mu \nu$SSM, accompanied by necessary numerical analysis, has been
discussed in ref. \cite{Ghosh-Roy}. In this work, neutrino masses and
mixing, consistent with the three flavour global neutrino data, are
reproduced even with the simplistic choice of flavour diagonal
neutrino Yukawa couplings $(Y^{ii}_\nu)$. Decay modes of the lightest
neutralino into two-body final states ($Z^0\nu_\ell$,~$W^{\pm}\ell$) 
have also been considered for various compositions of the lightest neutralino. 
In addition, correlations between neutrino mixing angles, and ratio of the decay
branching ratios into $W^{\pm}$-charged lepton are studied there as a
possible test of this model at the Large Hadron Collider (LHC).

Among other related studies of importance, neutrino mass generation and collider 
aspects of this model, with one and two generation(s) of the right handed neutrinos, 
have been addressed in ref.\cite{Porod-Bartl}. For one right handed neutrino, 
neutrino mass generation has been studied there upto one-loop level. Decays of the
lightest neutralino into all possible final states are also studied in this
reference. Constraints on complex vacuum expectation values (VEVs) from
electroweak symmetry breaking (EWSB), and its consequence on the neutrino
sector are studied in ref.\cite{munoz-lopez-3}, where issues concerning spontaneous
CP-violation are also addressed. The role of gravitino as a dark matter candidate 
in this model was studied in ref.\cite{munoz-lopez-4}. This paper also highlighted 
the prospects of detecting gamma rays from decaying gravitinos. For an overview of
various other aspects of $\mu\nu$SSM, see the recent review \cite{munoz-munussm}.

In this work, we study in detail the effect of radiative corrections, upto
one-loop, to the neutrino masses and mixing, consistent with the three flavour
global data \cite{boris_review_08, Schwetz-Valle}. As mentioned before, a
similar study, but with just one generation of right handed neutrino, was
carried out in \cite{Porod-Bartl}. However, a full study addressing both
neutrino masses and the bilarge mixing pattern, with a complete set of 
one-loop corrections with all three generations of left and right handed 
neutrinos, has so far been lacking. This is exactly what we attempt in the present 
work. We also perform a systematic study to identify the crucial parameters of the 
model, which control the tree level or the one-loop dominance in the neutrino sector.

As shown in Ref.\cite{Ghosh-Roy,Porod-Bartl}, a very attractive feature of 
this model is that the ratios of certain decay branching ratios show very nice correlation
with the neutrino mixing angles. This is very similar to bilinear R-parity violating
models \cite{Roy-Mukhopadhyaya-Vissani,choi-chun-kang-lee,porod-hirsch-romao-valle}. 
Nevertheless, one should note certain differences in these two cases. In $\mu\nu$SSM 
lepton number is broken explicitly in the superpotential by terms which are trilinear 
as well as linear in singlet neutrino superfields. In addition to that there are 
lepton number conserving terms involving the singlet neutrino superfields with dimensionless neutrino 
Yukawa couplings. After the electroweak symmetry breaking these terms can generate the effective bilinear 
R-parity violating terms as well as the $\Delta L$ =2 Majorana mass terms for the singlet neutrinos in the 
superpotential. In general, there are corresponding soft supersymmetry breaking terms in the scalar potential.
Thus the parameter space of this model is much larger compared to the bilinear R-parity violating model. 
Hence, in general, one would not expect a very tight correlation between the neutrino mixing angles and the
ratios of decay branching ratios of the LSP. However, under certain simplifying assumptions (as discussed 
in Sec. 3.2), one can reduce the number of free parameters and in those cases it is possible that the above 
correlations reappear. As mentioned earlier, this has been studied in great detail for the two body $\ell-W$ 
final states in \cite{Ghosh-Roy} and for all possible two and three body final states in \cite{Porod-Bartl}. 
Let us note in passing that such a nice correlation is lost in the general scenario of bilinear-plus-trilinear 
R-parity violation \cite{choi-chun-kang-lee}.      

Another important difference between $\mu\nu$SSM and the bilinear R-parity violating model in the context of 
the decay of the LSP (assumed to be the lightest neutralino in this case) is that in $\mu\nu$SSM the lightest 
neutralino can have a significant singlet neutrino ($\nu^c$) contribution. In this case, the correlation between 
neutrino mixing angles and decay branching ratios of the LSP is different \cite{Ghosh-Roy,Porod-Bartl} compared 
to the cases when the dominant component of the LSP is either a bino, or a higgsino or a Wino. This gives us a
possibility of distinguishing between different R-parity violating models through the observation of the
decay branching ratios of the LSP in collider experiments \cite{Ghosh-Roy,Porod-Bartl}. In addition, the decay of 
the lightest neutralino will show displaced vertices in collider experiments and when the lightest neutralino
is predominantly a singlet neutrino, the decay length can be of the order of several meters for a lightest 
neutralino mass in the neighbourhood of 50 GeV \cite{Porod-Bartl}. This is very different from the bilinear R-parity 
violating model where for a Bino LSP of similar mass the decay length is less than or of the order of a meter 
or so \cite{porod-hirsch-romao-valle}. 

The paper is organized as follows. We start with a brief introduction to the
model in section \ref{Electroweak symmetry breaking} and discuss the electroweak symmetry breaking
conditions. The neutrino sector is discussed in section \ref{The neutrinos sector} in details,
accompanied with necessary analytical results. We discuss the observed pattern
of neutrino masses and mixing upto one-loop corrections. We present a
comprehensive discussion on the results of our numerical analysis of neutrino
masses and mixing in section \ref{Numerical results of neutrino mass and mixing}. The three broad scenarios, namely, normal 
hierarchy, inverted hierarchy, and quasi-degenerate neutrinos, are taken up in 
turn in this section. We conclude in section \ref{Summary and conclusion}. Various technical 
details, such as different mass matrices, couplings, Feynman
rules and the expressions for one-loop contributions are relegated to the 
appendices.


\section{Electroweak symmetry breaking in $\mu \nu$SSM}
\label{Electroweak symmetry breaking}
The superpotential for $\mu \nu$SSM includes three gauge-singlet right
handed neutrino superfields (${\hat \nu}_i^c$~($i = e, {\mu},
{\tau}$)) along with the usual MSSM superfields. The superpotential of
$\mu \nu$SSM along the lines of ref.\cite{munoz-lopez_fogliani} is
\bea 
W &=& \epsilon_{ab}(Y^{ij}_u\hat H^b_2\hat Q^a_i\hat u^c_j +
Y^{ij}_d\hat H^a_1 \hat Q^b_i\hat d^c_j + Y^{ij}_e\hat H^a_1\hat
L^b_i\hat e^c_j + Y^{ij}_\nu
\hat H^b_2\hat L^a_i\hat \nu^c_j)\nonumber \\
&-&\epsilon_{ab} \lambda^i\hat \nu^c_i\hat H^a_1\hat H^b_2 +
\frac{1}{3}\kappa^{ijk}\hat \nu^c_i\hat \nu^c_j\hat \nu^c_k,
\label{superpotential}
\eea
where $\hat H_1$ and $\hat H_2$ are the Higgs
superfields that have Yukawa couplings with down- and up-type quarks, 
respectively. $\hat Q_i$ are doublet quark superfields,
${\hat u}^c_j$ (${\hat d}^c_j$) are singlet up-type (down-type) quark
superfields, $\hat L_i$ are doublet lepton superfields, and ${\hat
  e}^c_j$ are singlet charged lepton superfields. The absence of any
bilinear terms in the superpotential is ensured by imposing a
$Z_3$ symmetry (which is also used in case of NMSSM). The effective
$\mu$-term is given by $\mu = \sum{\lambda^i}{v^c_i}$, where ${v^c_i}$
is the VEV obtained by the `$i$'-th right handed sneutrino (scalar
component of $\hat \nu^c_i$) after EWSB. The characteristic bilinear $R_p$ 
violating terms ($\varepsilon_{i} {\hat L}_i {\hat H}_2$) appear in a similar way 
after the EWSB. These terms are given by $\varepsilon^i = \sum Y^{ij}_\nu
v^c_j$. The last two terms of the superpotential (see eq.(\ref{superpotential})) 
violate $R_p$ through L-violation. The last term, with the coefficient $\kappa^{ijk}$, is 
included in order to avoid an unacceptable axion associated to the breaking of a
global $\rm U(1)$ symmetry \cite{ellis-gunion-haber}. This term
generates effective Majorana masses for the singlet neutrinos at the
electroweak scale.

It has been shown earlier \cite{munoz-lopez_fogliani,munoz-lopez-2,Ghosh-Roy} that the above
$R_P$-violating superpotential (see eq.(\ref{superpotential})) provides the
minimal structure, sufficient for both generating a neutrino mass pattern, and
offering a solution to the $\mu$-problem. There have been studies
\cite{Mukhopadhyaya-Srikanth-prd,Chang-Gouvea} on $R_P$-violating scenarios including right
handed neutrinos, which use a subset of this minimal superpotential, but
without any attempt to address the $\mu$-problem. Similar remarks apply to
earlier works \cite{Farzan-Valle} aimed at explaining the baryon asymmetry of
the universe through leptogenesis, using the term ${\hat \nu}^c {\hat H}_1
{\hat H}_2$.

The frequently discussed trilinear $L$-violating terms, driven by the
well-known $\lambda$- and $\lambda^{\prime}$-type couplings, need not appear
explicitly in this model. The reason is the following; the above
superpotential, as has already been stated, can lead to `bilinear' terms of
the form $L_i H_2$ once the right sneutrinos acquire VEV, and, once such terms
arise, they can effectively lead to the $\lambda$ and $\lambda^{\prime}$-type
terms \cite{Roy-Mukhopadhyaya,diaz-romao-valle}. Here, as a digression, let us mention that the
spontaneous breakdown of the $Z_3$ symmetry through right-sneutrino VEV can in
general lead to the formulation of domain walls \cite{domain-wall}. The
associated problems can, however, be ameliorated through well-known methods
\cite{domain-wall-soln}.



Coming back to $\mu\nu$SSM, if we confine ourselves to the framework
of supergravity mediated supersymmetry breaking, the Lagrangian
$\mathcal{L}_{\text{soft}}$, containing the
soft-supersymmetry-breaking terms is given by
\bea
-\mathcal{L}_{\text{soft}} &=&
(m_{\widetilde{Q}}^2)^{ij} {\widetilde Q^{a^*}_i} \widetilde{Q^a_j}
+(m_{\widetilde u^c}^{2})^{ij}
{\widetilde u^{c^*}_i} \widetilde u^c_j
+(m_{\widetilde d^c}^2)^{ij}{\widetilde d^{c^*}_i}\widetilde d^c_j
+(m_{\widetilde{L}}^2)^{ij} {\widetilde L^{a^*}_i}\widetilde{L^a_j} \nonumber \\
&+&(m_{\widetilde e^c}^2)^{ij}{\widetilde e^{c^*}_i}\widetilde e^c_j 
+ m_{H_1}^2 {H^{a^*}_1} H^a_1 + m_{H_2}^2 {H^{a^*}_2} H^a_2 +
(m_{\widetilde{\nu}^c}^2)^{ij}  {\widetilde{\nu}^{c^*}_i} \widetilde\nu^c_j \nonumber \\
&+& \epsilon_{ab} \left[
(A_uY_u)^{ij} H_2^b\widetilde Q^a_i \widetilde u_j^c +
(A_dY_d)^{ij} H_1^a \widetilde Q^b_i \widetilde d_j^c +
(A_eY_e)^{ij} H_1^a \widetilde L^b_i \widetilde e_j^c + \text{H.c.}  \right] 
\nonumber \\
&+&\left[\epsilon_{ab}(A_{\nu}Y_{\nu})^{ij} H_2^b \widetilde L^a_i \widetilde 
\nu^c_j-\epsilon_{ab} (A_{\lambda}\lambda)^{i} \widetilde \nu^c_i H_1^a  H_2^b+
\frac{1}{3} (A_{\kappa}\kappa)^{ijk} \widetilde \nu^c_i \widetilde \nu^c_j \widetilde 
\nu^c_k\ + \text{H.c.} \right] \nonumber \\
&-& \frac{1}{2} \left(M_3 \widetilde{\lambda}_3 \widetilde{\lambda}_3
+ M_2 \widetilde{\lambda}_2 \widetilde{\lambda}_2 + M_1 \widetilde{\lambda}_1 
\widetilde{\lambda}_1 + \text{H.c.} \right).
\label{Lsoft}
\eea
The first two lines of eq.(\ref{Lsoft}) consist of squared-mass terms of
squarks, sleptons and Higgses, the next two lines contain the trilinear scalar
couplings, while in the last line, $M_3, M_2$, and $M_1$ represent the
Majorana masses corresponding to $SU(3)$, $SU(2)$ and $U(1)$ gauginos
$\widetilde{\lambda}_3, \widetilde{\lambda}_2$, and $\widetilde{\lambda}_1$,
respectively. The tree-level scalar potential receives the usual D and F term
contributions, in addition to the terms from $\mathcal{L}_{\text{soft}}$.


We adhere to the $CP$-preserving case, so that only the real parts of
the neutral scalar fields develop, in general, the following VEVs,
\bea
\langle H_1^0 \rangle = v_1 \, , \quad \langle H_2^0 \rangle = v_2 \,
, \quad \langle \widetilde \nu_i \rangle = v^{\prime}_i \, , \quad
\langle \widetilde \nu_i^c \rangle = v_i^c.
\label{vevs}
\eea
The tree level neutral scalar potential looks like 
\bea \langle V_{\text{neutral}}\rangle &=&
\left|\sum_{i,j}Y^{ij}_{\nu}{v^{\prime}_i}{v^c_j}-
  \sum_{i}\lambda^i{v^c_i} v_1\right|^2\ + \sum_{j}\left|\sum_{i}
  Y^{ij}_{\nu}{v^{\prime}_i}v_2
  -\lambda^jv_1v_2+\sum_{i,k} \kappa^{ijk}{v^c_i}{v^c_k}\right|^2 \nonumber \\
&+& \left|\sum_{i}\lambda^i{v^c_i}v_2\right|^2 +
\sum_{i}\left|\sum_{j}Y^{ij}_{\nu}v_2 {v^c_j}\right|^2
+(\frac{g_1^2+g_2^2}{8}) \left[\sum_{i}|v'_i|^2+|v_1|^2 -
  |v_2|^2\right]^2 \nonumber \\
&+& \left [\sum_{i,j}(A_\nu Y_\nu )^{ij} {v'_i} {v^c_j}v_2 -\sum_{i}
  (A_\lambda \lambda )^{i} {v^c_i}v_1v_2 +
  \sum_{i,j,k}\frac{1}{3}(A_\kappa
  \kappa )^{ijk}v^c_iv^c_jv^c_k + {\rm H.c.} \right]\nonumber \\
&+& \sum_{i,j} (m_{\widetilde{L}}^2)^{ij}{v'_i}^* {v'_j}
+\sum_{i,j}(m_{\widetilde{\nu}^c}^2)^{ij} {v^{c^*}_i}{v^c_j} +
m_{H_2}^2|v_2|^2 + m_{H_1}^2|v_1|^2 .
\label{Vneut}
\eea
It is important to notice that the potential is bounded from below as the
coefficients of the fourth power of all the eight superfields are positive. We
further assume that all the parameters present in the scalar potential are
real. From eq.(\ref{Vneut}), the minimization conditions in terms of
$v^c_i~v'_i,v_2,~v_1$ can be derived (the equations are provided in appendix
{\ref{Minimization equations}}). The minimization conditions for $\mu \nu$SSM
have also been addressed in \cite{munoz-lopez-2,Ghosh-Roy,Porod-Bartl}
. Similar conditions, but for complex VEVs, have been discussed in
ref.\cite{munoz-lopez-3}. Note that in order to generate correct order of magnitudes 
for the light neutrino masses through the TeV scale seesaw mechanism, one requires 
smaller values for neutrino Yukawa couplings $(Y^{ij}_\nu$ $\sim$  $\cal{O}$ $(10^{-6}))$ 
and left handed sneutrino VEVs $(v^\prime_i$ $\sim$  $\cal{O}$ $(10^{-4})$ ${\rm{GeV}}$).

\section{The neutrino sector}
\label{The neutrinos sector}

\subsection{Neutral fermions}
\label{Neutral fermions}
In this model, three $SU(2)_L$ doublet neutrinos ($\nu_i$) and three
gauge-singlet right handed neutrinos ($\nu^c_i$) mix with the MSSM neutralinos
(two neutral gauginos and two neutral higgsinos) due to L-violating
interactions (see eq.(\ref{superpotential})). The resulting neutralino mass
matrix therefore is of dimension $10\times10$. The mixing among various
current eigenstates are governed by the VEVs of various neutral scalar fields 
(namely, $H_1^0, H_2^0, \widetilde \nu_i, \widetilde \nu^c_i$). This matrix has been
addressed in refs. \cite{munoz-lopez_fogliani,munoz-lopez-2,Ghosh-Roy,Porod-Bartl} for
real VEVs, and in ref. \cite{munoz-lopez-3} for complex VEVs.

In the weak interaction basis, defined by,
\beq
\label{neutralino_basis}
{\Psi^0}^T = \left(\widetilde B^0, \widetilde W_3^0, \widetilde H_1^0, 
\widetilde H_2^0,{\nu^c_e},{\nu^c_{\mu}},{\nu^c_{\tau}},{\nu_e},{\nu_{\mu}},
{\nu_{\tau}} \right),
\eeq
the neutral fermion mass term in the Lagrangian is of the form
\beq
\label{weak-basis-Lagrangian-neutralino}
{\mathcal{L}_{neutral}^{mass}} = -\frac{1}{2}{{\Psi^0}^T} \mathcal{M}_n 
{\Psi^0} + \text{H.c.},
\eeq
where $\mathcal{M}_n$ is the $10\times10$ modified neutralino mass
matrix, and is given by
\beq 
\mathcal{M}_n =
\left(\begin{array}{cc}
    M_{7\times 7} & m_{3\times 7}^T \\
    m_{3\times 7} & 0_{3\times 3}
\end{array}\right).
\label{neutralino-seesaw}
\eeq
Here, using eq.(\ref{Abbrevations}),
\beq
M_{7\times7} =
\left(\begin{array}{ccccccc}
M_1 & 0 & -\frac{g_1}{\sqrt{2}}v_1 & \frac{g_1}{\sqrt{2}}v_2 & 0 & 0 & 0 \\ \\
0 & M_2 & \frac{g_2}{\sqrt{2}}v_1 & -\frac{g_2}{\sqrt{2}}v_2 & 0 & 0 & 0 \\ \\
-\frac{g_1}{\sqrt{2}}v_1 & \frac{g_2}{\sqrt{2}}v_1 & 0 & -{\mu} & 
-{\lambda^e}v_2 & -{\lambda^{\mu}}v_2 & -{\lambda^{\tau}}v_2 \\ \\
\frac{g_1}{\sqrt{2}}v_2 & -\frac{g_2}{\sqrt{2}}v_2 & -{\mu} & 0 & {\rho^e} 
& {\rho^{\mu}} & {\rho^{\tau}}\\ \\
0 & 0 & -{\lambda^e}v_2 & {\rho^e} & 2 {u^{ee}_c} & 2 {u^{e{\mu}}_c} & 
2 {u^{e{\tau}}_c}\\ \\
0 & 0 & -{\lambda^{\mu}}v_2 & {\rho^{\mu}} & 2{u^{{\mu}e}_c} & 
2 {u^{{\mu}{\mu}}_c} & 2 {u^{{\mu}{\tau}}_c}\\ \\
0 & 0 & -{\lambda^{\tau}}v_2 & {\rho^{\tau}} & 2 {u^{{\tau}e}_c} & 
2{u^{{\tau}{\mu}}_c} & 2 {u^{{\tau}{\tau}}_c}
\end{array}\right),
\label{neutralino_7x7}
\eeq
and
\beq
m_{3\times7} =
\left(\begin{array}{ccccccc}
-\frac{g_1}{\sqrt{2}}{v'_e} & \frac{g_2}{\sqrt{2}}{v'_e} & 0 & {r^e_c} & 
Y_{\nu}^{ee} v_2 & Y_{\nu}^{e{\mu}} v_2 & Y_{\nu}^{e{\tau}} v_2\\ \\
-\frac{g_1}{\sqrt{2}}{v'_{\mu}} & \frac{g_2}{\sqrt{2}}{v'_{\mu}} & 0 & 
{r^{\mu}_c} & Y_{\nu}^{{\mu}e} v_2 & Y_{\nu}^{{\mu}{\mu}} v_2 & 
Y_{\nu}^{{\mu}{\tau}} v_2\\ \\
-\frac{g_1}{\sqrt{2}}{v'_{\tau}} & \frac{g_2}{\sqrt{2}}{v'_{\tau}} & 0 &
{r^{\tau}_c}  & Y_{\nu}^{{\tau}e} v_2 & Y_{\nu}^{{\tau}{\mu}} v_2 & 
Y_{\nu}^{{\tau}{\tau}} v_2
\end{array}\right).
\label{neutralino_3x7}
\eeq
The matrix $M_{7\times7}$ contains the $4\times4$ block (upper left) of MSSM
neutralinos as well as a $3\times3$ block (bottom right) of gauge-singlet
neutrinos and mixing terms between them. The null $3\times3$ block in
$\mathcal{M}_n$ signifies the absence of Majorana mass terms for the left
handed neutrinos. The elements of $m_{3\times7}$ contain either left handed
sneutrino VEVs $(v^{\prime}_i)$ or Higgs VEVs multiplied by neutrino Yukawa 
couplings $(Y_{\nu}^{ij})$, and hence, are of much smaller magnitudes compared 
to the entries of $M_{7\times7}$. This feature ensures a {\it{seesaw}}-like 
structure of $\mathcal{M}_n$.

This $10\times10$ symmetric matrix $\mathcal{M}_n$ can be diagonalized
with a $10\times10$ unitary matrix $N$ to obtain the physical
neutralino states. The mass eigenstates are defined by,
\beq
\label{neutralino_mass_eigenstate}
{\widetilde \chi}^0_i= N_{ij} \Psi^0_j, \quad i,j=1,...,10,
\eeq
where $N$ satisfies
\beq
\label{neutralino_mass_eigenstate_matrixform}
N^* \mathcal{M}_n N^{-1} = \mathcal{M}^0_D,
\eeq
with the diagonal neutralino mass matrix denoted as $\mathcal{M}^0_D$.
Seven eigenvalues of this matrix turn out to be heavy, i.e. of the
order of the electroweak scale, and thus correspond to the physical
neutralinos. The remaining three light eigenvalues correspond to the
masses of three SM-neutrinos. One can therefore write
eq.(\ref{neutralino_mass_eigenstate_matrixform}) alternatively as
\beq
\label{neutralino_mass_eigenstate_matrixform-1}
N^* \mathcal{M}_n N^{-1} = \rm{diag}(m_{{\widetilde \chi}^0_i},m_j),
\eeq
where $i=1,...,7$ and $j=1,2,3$.

Assuming small ${{R}}_P$ violation, it is possible to carry out a
perturbative diagonalization of the $10\times10$ neutralino mass
matrix (see \cite{Schechter-Valle}), by defining \cite{Hirsch-Valle}
a matrix $\xi$ as
\beq
\label{expansion-parameter}
\xi=m_{3\times7}.M^{-1}_{7\times7}.
\eeq
If the elements of $\xi$ satisfy $\xi_{ij} \ll 1$, then this can be
used as an expansion parameter to get an approximate analytical
solution for the matrix $N$ (see
eq.(\ref{neutralino_mass_eigenstate_matrixform})). A general expression for the 
elements of $\xi$ with simplified assumptions can be written in the form 
$\mathcal{A}a_i + \mathcal{B}b_i + \mathcal{C}c_i$,
where
\beq 
a_i = Y_{\nu}^{ii} v_2, ~c_i = {v'_i}, ~b_i = (Y_{\nu}^{ii} v_1 + 3 {\lambda}
{v'_i}) = (a_i\cot\beta + 3 \lambda c_i),
\label{specifications}
\eeq
with ${i} = {e,\mu,\tau} ~\equiv{1,2,3}$, $\tan\beta = \frac{v_2}{v_1}$ and
$\mathcal{A},\mathcal{B},\mathcal{C}$ are complicated functions of various
parameters of the model. The complete expressions for the elements of $\xi$ are given in appendix
\ref{Details of expansion matrix}. Here we neglect the subdominant terms $\cal{O}$ $\sim$ 
$\frac{v'^3}{{\tilde m}^3}$, $\frac{Y_\nu v'^2}{{\tilde m}^2}$,
$\frac{Y_\nu^2 v'}{{\tilde m}}$, where $\tilde m$ is the electroweak (or 
supersymmetry breaking) scale. 

The mixing matrix $N$ in leading order in $\xi$ is given by
\beq 
N^* = \left(\begin{array}{cc}
    \mathcal{N}^* & 0 \\
    0 & U^T_{\nu}
\end{array}\right)
\left(\begin{array}{cc}
1-\frac{1}{2} \xi^{\dagger} \xi & \xi^{\dagger} \\
-\xi & 1-\frac{1}{2} \xi \xi^{\dagger}
\end{array}\right).
\label{neutralino-mixing-matrix}
\eeq
The $10\times10$ neutralino mass matrix $\mathcal{M}_n$ can approximately be
block-diagonalized to the form {\it{diag($M_{7\times7},{M^{seesaw}_{\nu}}$)}},
by the matrix defined in eq.(\ref{neutralino-mixing-matrix}). The matrices
$\mathcal{N}$ and $U_{\nu}$, defined in eq.(\ref{neutralino-mixing-matrix}),
are used to diagonalize ${M_{7\times7}}$ and ${M^{seesaw}_{\nu}}$ in the following manner,
%
\bea
\label{diag-matrix}
& &\mathcal{N}^* M_{7\times7} \mathcal{N}^{\dagger} = 
\rm{diag} (m_{{\widetilde \chi}^0_i}), \nonumber \\ 
& &\mathcal{U}^T_\nu {M^{seesaw}_{\nu}} \mathcal{U}_\nu = 
\rm{diag} (m_1,m_2,m_3).\nonumber \\
\eea
%
\subsection{Seesaw mechanism and tree level neutrino mass} 

The effective light neutrino mass matrix ${M^{seesaw}_{\nu}}$, arising
via the seesaw mechanism in presence of explicit lepton number
violation, is in general given by
\beq 
{M^{seesaw}_{\nu}} = -{m_{3\times7}} {M_{7\times7}^{-1}}
{m_{3\times7}^T}.
\label{seesaw_formula}
\eeq
The eigenvalues and eigenvectors of the $3\times3$ matrix
${M^{seesaw}_{\nu}}$ were computed in ref.\cite{Ghosh-Roy} with the
simplifying assumption of a flavour diagonal structure of the neutrino
Yukawa couplings $Y^{ij}_\nu$. Three flavour global neutrino data were
fitted with this assumption, and it was observed that all three
neutrinos acquire masses even at the tree level.

An approximate analytical expression for the elements of
${M^{seesaw}_{\nu}}$ at tree level, as obtained in
ref.\cite{Ghosh-Roy}, with certain simplifying assumptions is given by
\beq
({M^{seesaw}_{\nu}})_{ij} = {\frac{1}{6 \kappa {v^c}}} {a_{i}}
{a_{j}}(1-3\delta_{ij}) + {\frac{2 A {v^c}}{3 \Delta}} {b_{i}}
{b_{j}}.
\label{mnuij-compact1}
\eeq
One can further rewrite eq.(\ref{mnuij-compact1}) in an elucidate form
given by
\beq
({M^{seesaw}_{\nu}})_{ij} = f_1 a_i a_j + f_2 c_i c_j + f_3 (a_i c_j + a_j c_i),
\label{mnuij-compact-recasted}
\eeq
 where $a_i$ and $c_i$ are given by eq.(\ref{specifications}) and
\bea
f_1 &=& \frac{1}{6 \kappa v^c} (1-3\delta_{ij}) + \frac{2 A v^c {\rm{cot}}^2\beta}{3 \Delta}, \nonumber \\
f_2 &=& \frac{2 A \lambda \mu}{\Delta},~~f_3 = \frac{2 A \mu {\rm{cot}}\beta}{3 \Delta}, 
\label{specifications-3}
\eea
with
\bea
\mu &=& 3 \lambda v^c, ~~A = (\kappa {v^c}^2 + \lambda v_1 v_2), \nonumber \\
\Delta &=& \lambda^2 (v^2_1 + v^2_2)^2 + 4 \lambda \kappa v_1 v_2 {v^c}^2 - 4 \lambda A \mu M,  \nonumber \\
\frac{1}{M} &=& \frac{g^2_1}{M_1} + \frac{g^2_2}{M_1}.
\label{specifications-4}
\eea
The reason for recasting eq.(\ref{mnuij-compact1}) in terms of 
$a_i$ and $c_i$ becomes clear when we will discuss our numerical results 
in section \ref{Numerical results of neutrino mass and mixing}.
\noindent In $({M^{seesaw}_{\nu}})_{ij}$, we neglect the subdominant 
terms of the order of $\frac{Y_\nu v'^3}{{\tilde m}^2}$, 
$\frac{Y_\nu^2 v'^2}{\tilde m}$ and ${Y_\nu^3 v'}$. 
For the convenience of the reader, let us also
mention here that we choose $\lm^i$, $~(A_{\lm} \lm)^i$,
$~\kappa^{ijk}$, $~(A_{\kappa} \kappa)^{ijk}$ and all soft masses to
be {\it{flavour diagonal}} and {\it{flavour blind}}, The neutrino
Yukawa couplings ($Y_{\nu}^{ij}$) and the corresponding soft terms
$(A_{\nu} Y_\nu)^{ij}$ are, however, chosen to be {\it{flavour
    diagonal}}.

\subsection{One loop corrections to the self energies}
\label{One loop corrections to the self energies}
In the regime of renormalizable quantum field theories, stability of any tree
level analysis must be re-examined in the light of radiative
corrections. Following this prescription, the results of neutrino masses and
mixing will be more robust, once tree level analysis is further improved by
incorporating radiative corrections. The radiative corrections may have
sizable effect on the neutrino data at one-loop level. Thus, although all
three SM neutrinos acquire non-zero masses in the $\mu \nu$SSM even at the
tree level \cite{Ghosh-Roy}, it is interesting to investigate the fate of
those tree level masses and mixing when exposed to one-loop corrections. With
this in view, in this section we perform a systematic study of the neutrino
mass and mixing with all possible one-loop corrections both analytically and
numerically. In the subsequent sections, while showing the results of one-loop
corrections, we try to explain the deviations (which may or may not be
prominent) from the tree level analysis. The complete set of one-loop diagrams are shown 
in figure\ref{one-loop-diagrams}. Before going into the details, let us
discuss certain relevant issues of one-loop correction and renormalization for
the neutralino-neutrino sector. The most general one-loop contribution to the
unrenormalized neutralino-neutrino two-point function can be expressed as
\beq 
i {\bf \Sigma}^{ij}_{\n \n}(p) = i\{\ps \left[P_L
  {\Sigma^L_{ij}}(p^2) + P_R {\Sigma^R_{ij}}(p^2)\right] -\left[P_L
  {\Pi^L_{ij}}(p^2) + P_R {\Pi^R_{ij}}(p^2)\right]\},
\label{loopgen-unrenorm}
\eeq
where $P_L$ and $P_R$ are defined in eq.(\ref{P-L-P-R}), $i,~j~=~1,...,10$ and
$p$ is the external momentum. The unrenormalized self-energies $\Sigma$ and
$\Pi$ depend on the squared external momentum $(p^2)$. The generic self
energies $\Sigma^{L(R)}_{ij},~\Pi^{L(R)}_{ij}$ of the (Majorana) neutrino must
be symmetric in its indices $i,j$. The resulting one-loop corrected mass
matrix using dimensional reduction ($\ovl{DR}$) scheme \cite{Siegel-Capper} is
given by
\beq
\label{one-loop-corrected-mass}
(\mathcal{M}^{\rm{tree + 1-loop}}_{\n})^{ij} = {m}_{\n}(\mu_R)\delta^{ij} +
{\frac{1}{2}}\left(\widetilde \Pi^V_{ij}({{m^2_i}}) + \widetilde
  \Pi^V_{ij}({{m^2_j}}) - {m_{\n_i}}{\widetilde
    \Sigma^V_{ij}}({{m^2_i}}) - {m_{\n_j}}{\widetilde
    \Sigma^V_{ij}}({{m^2_j}})\right), 
\eeq
with
\bea
\label{Sigma-Pi-renomalized}
\widetilde \Sigma^V_{ij} &=& \frac{1}{2} (\widetilde \Sigma^L_{ij} + \widetilde
\Sigma^R_{ij}),
\nonumber \\
\widetilde \Pi^V_{ij} &=& \frac{1}{2} (\widetilde \Pi^L_{ij} + \widetilde
\Pi^R_{ij}). 
 \eea
where the tree level neutralino mass $({m}_{\n})$ is defined at the
renormalization scale $\mu_R$, set at the electroweak scale. Here, the 
word {\it{neutralino mass}} stands for all the {\it{ten}} eigenvalues 
of the $10\times10$ neutralino mass matrix. The self-energies $\Sigma,~\Pi$ 
are also renormalized in the $\ovl{DR}$ scheme \cite{Siegel-Capper} and denoted
by $\widetilde \Sigma$ and $\widetilde \Pi$ respectively. The detailed
expressions of $\widetilde \Sigma^V_{ij}$ and $\widetilde \Pi^V_{ij}$
are given in appendix \ref{self-energy-sigma-pi}.
%


\subsection{Radiative corrections to neutrino mass terms}
\label{Radiative corrections to neutrino mass}
In this section we consider the effect of radiative corrections to the light
neutrino masses. Let us recapitulate some of the earlier work regarding
one-loop corrections to the neutralino-neutrino sector. The complete set of
radiative corrections to the neutralino mass matrix in the $R_P$ conserving
MSSM was discussed in ref.\cite{Pierce-Papadopoulos}, and the leading order
neutrino masses has been derived in ref.\cite{Hall-Suzuki}. One-loop radiative
corrections to the neutrino-neutralino mass matrix in the context of a
$R_P$-violating model were calculated in ref.\cite{Hempfling-Rpv} using
't-Hooft-Feynman gauge. In ref.\cite{Hirsch-loop-FD}, $R_{\xi}$ gauge has been
used to compute the corrections to the neutrino-neutralino mass matrix at
one-loop level in an $R_p$-violating scenario. Neutrino mass
generation at the one-loop level in other variants of $R_P$-violating MSSM has
also been addressed in
refs. \cite{Grossman-Haber-prd,Grossman-Haber,Grossman-Rakshit-prd,Davidson-Losada-jhep,Davidson-Losada-prd,Abada-Davidson-Losada,Abada-Bhattacharyya-Losada,Davidson-Losada-Rius,Borzumati-Lee,Chun-Jung-Park,Dedes-Rimmer-Rosiek,Mukhopadhyaya-Srikanth-prd,Dedes-Haber-Rosiek}.

We begin by outlining the strategy of our analysis. We
start with a general $10\times10$ neutralino matrix, with off-diagonal entries
as well, which has a {\it{seesaw structure}} in the {\it{flavour-basis}} (see
eq.(\ref{neutralino-seesaw})). Schematically, we can rewrite
eq.(\ref{neutralino-seesaw}) as,
\beq
\mathcal{M}_n = \left(\begin{array}{cc}
    M_f & m^T_{D_f} \\
    m_{D_f} & 0
\end{array}\right),
\label{neutralino-seesaw-schematic}
\eeq
where the orders of the block matrices are as those indicated in
eq.(\ref{neutralino-seesaw}), and the subscript `$f$' denotes the flavour
basis. Here $M_f$ stands for the $7\times7$ Majorana mass matrix of the heavy
states, while $m_{D_f}$ contains the $3\times7$ Dirac type masses for the left
handed neutrinos. In the next step, instead of utilising the seesaw structure
of this matrix to generate the effective light neutrino mass matrix for the
three active light neutrino species, we {\it{diagonalize}} the entire
$10\times10$ matrix $\mathcal{M}_n$. The diagonal $10\times10$ matrix
$M^{0}_D$ thus contains tree level neutralino masses, which we symbolically
write as
\beq
\mathcal{M}_D^{0} = \left(\begin{array}{cc}
    M_m & 0 \\
    0 & m_m
\end{array}\right),
\label{neutralino-tree-level-schematic}
\eeq
where $M_m~(m_m)$ are the masses of the heavy states (left handed
neutrinos). At this stage we turn on all possible one-loop interactions, so
that the $10\times10$ matrix $\mathcal{M}_D^{0}$ picks up radiatively
generated entries, both diagonal and off-diagonal. The resulting one-loop
corrected Lagrangian for the neutralino mass terms, following
eq.(\ref{weak-basis-Lagrangian-neutralino}), can be written as
\beq
\label{loop-correcetd-Lagrangian-neutralino-a} {\mathcal{L}^{\prime}}
= -\frac{1}{2}{{\widetilde \chi^{0^T}}} \left(\mathcal{M}^0_D +
  \mathcal{M}^1 \right) {\widetilde \chi^0} + \text{H.c.}, 
\eeq
where $\mathcal{M}^1$ contains the effect of one-loop corrections. The
$10\times10$ matrix $\mathcal{M}^0_D$ is diagonal, but the matrix
$\mathcal{M}^1$ is a general symmetric matrix with off diagonal
entries.

One can rewrite the above equation, using
eqs.(\ref{neutralino_mass_eigenstate} and
\ref{neutralino_mass_eigenstate_matrixform}), as
\beq
\label{loop-correcetd-Lagrangian-neutralino-b}
{\mathcal{L}^{\prime}} = -\frac{1}{2}{{\Psi^0}^T} \left(\mathcal{M}_n
  + N^T \mathcal{M}^1 N\right) {\Psi^0} + \text{H.c.}.  
\eeq
This is nothing but the {\it{one-loop corrected}} neutralino mass term in the Lagrangian in the
flavour basis. Symbolically,
\beq
\label{loop-correcetd-Lagrangian-neutralino-c}
{\mathcal{L}^{\prime}} = -\frac{1}{2}{{\Psi^0}^T} 
\mathcal{M}^{\prime} {\Psi^0} + \text{H.c.},
\eeq
with the $10\times10$ matrix $\mathcal{M}^{\prime}$ having the form
\beq
\mathcal{M}^{\prime} = 
\left(\begin{array}{cc}
M_f + \Delta{M}_f & (m_{D_f} + \Delta{m_{D_f}})^T \\
m_{D_f} + \Delta{m_{D_f}} & \Delta{m}_f
\end{array}\right).
\label{neutralino-tree-plus-one-loop-level-schematic}
\eeq
The quantities $\Delta{M}_f$ and $\Delta{m}_f$ stand for one-loop corrections
to the heavy neutralino states and light neutrino states respectively, in the
flavour basis $\Psi^0$. The entity $\Delta{m_{D_f}}$ arises because of the off
diagonal interactions, i.e. between the heavy neutralinos and the light
neutrinos, in the same basis. Note that all of $\Delta M_f$, $\Delta m_{D_f}$,
$\Delta m_f$ in the $\chi_0$ basis are given by the second term on the right
hand side of eq.(\ref{one-loop-corrected-mass}). We suitably transform them
into the basis $\Psi^0$ with the help of neutralino mixing matrix
$N$. Interestingly, the matrix $\mathcal{M}^{\prime}$ once again possesses a
seesaw structure, and one can therefore write down the one-loop corrected
effective light neutrino mass matrix as
\beq ({M}^{\nu^{\prime}})_{\rm{eff}} \approx
\Delta{m}_f - (m_{D_f} + \Delta{m_{D_f}})(M_f +
\Delta{M}_f)^{-1}((m_{D_f} + \Delta{m_{D_f}})^T).
\label{mass-basis-seesaw-schematic}
\eeq
Let us now present an approximate form of
eq.(\ref{mass-basis-seesaw-schematic}). For simplicity, let us begin by
assuming the quantities present in eq.(\ref{mass-basis-seesaw-schematic}) to
be c-numbers (not matrices). In addition, assume $M_f \gg \Delta{M}_f$
(justified later), so that eq.(\ref{mass-basis-seesaw-schematic}) may be
written as,
\beq ({M}^{\nu^{\prime}})_{\rm{eff}} \approx \Delta{m}_f - \delta \times
{M_f}\left\{\left(\frac{m_{D_f}}{M_f}\right)^2 + 2
  \left(\frac{m_{D_f}}{M_f}\right) \left(\frac{\Delta
      m_{D_f}}{M_f}\right) + \left(\frac{\Delta m_{D_f}}{M_f}\right)^2
\right\},
\label{mass-basis-seesaw-schematic-2}
\eeq
with $\delta = \left(1-\frac{\Delta M_f}{M_f}\right)$. Now, even when $\Delta
m_{D_f}$ $\sim$ $\frac{1}{16 \pi^2}$ $m_{D_f}$ and $\Delta
M_f$ $\sim$ $\frac{1}{16 \pi^2}$ $M_f$, eq.(\ref{mass-basis-seesaw-schematic-2}) looks like

\beq 
({M}^{\nu^{\prime}})_{\rm{eff}} \approx \Delta{m}_f - {M_f}
\left(1-\frac{1}{16 \pi^2}\right)\left\{\left(\frac{m_{D_f}}{M_f}\right)^2 +
\frac{2}{16 \pi^2} \left(\frac{m_{D_f}}{M_f}\right)^2 + \frac{1}{256 \pi^4}
\left(\frac{m_{D_f}}{M_f}\right)^2 \right\}.
\label{mass-basis-seesaw-schematic-3}
\eeq
Thus, up to a very good approximation one can rewrite
eq.(\ref{mass-basis-seesaw-schematic-3}) as
\beq
({M}^{\nu^{\prime}})_{\rm{eff}} \approx
\Delta{m}_f - {M_f} \left(\frac{m_{D_f}}{M_f}\right)^2.
\label{mass-basis-seesaw-schematic-4}
\eeq
Reimposing the matrix structure and using eq.(\ref{seesaw_formula}),
eq.(\ref{mass-basis-seesaw-schematic-4}) can be modified as,
\beq
({M}^{\nu^{\prime}})_{\rm{eff}} \approx
\Delta{m}_f + {M^{seesaw}_{\nu}}.
\label{mass-basis-seesaw-schematic-5}
\eeq
The eigenvalues of the $3\times3$ one-loop corrected neutrino mass matrix
$({M}^{\nu^{\prime}})_{\rm{eff}}$ thus correspond to one-loop corrected light
neutrino masses. In conclusion, it is legitimate to calculate one-loop
corrections to the $3\times3$ light neutrino mass matrix only, and diagonalize
it to get the corresponding one-loop corrected mass eigenvalues. 

Let us denote the one-loop corrections to the masses of heavy neutralinos and
light neutrinos in the basis $\widetilde \chi^0$ by $\Delta M$ and $\Delta m$
respectively. The one-loop corrections arising from neutralino-neutrino
interactions is denoted by $\Delta m_D$ in the same basis. The tree level
neutralino mixing matrix $N$ can then be written as,
\beq
N =
\left(\begin{array}{cc}
\widetilde N_{7\times 7} & \widetilde N_{7\times 3} \\
\widetilde N_{3\times 7} & \widetilde N_{3\times 3}
\end{array}\right),
\label{neutralino-mixing-matrix-block-form}
\eeq
where the entries of the matrices $\widetilde N_{7\times
  3},~\widetilde N_{3\times 7}$ are $\sim \cal{O}$
$({m_D^{\nu}}/{M_{\widetilde \chi^0}})$, due to very small
neutrino-neutralino mixing \cite{Han-Atre}. The quantities $m_D^{\nu}$
and $M_{\widetilde \chi^0}$ represent the Dirac mass of neutrino and
the Majorana mass of neutralino. From
eq.(\ref{loop-correcetd-Lagrangian-neutralino-b}), it is easy to
figure out the relation between $\Delta m$ and $\Delta m_f$ as,
\beq 
\Delta m_{f} = {\widetilde N_{7\times 3}^T} {\Delta M} {\widetilde
  N_{7\times 3}} + {\widetilde N_{7\times 3}^T} {\Delta m_D^T} {\widetilde
  N_{3\times 3}} + {\widetilde N_{3\times 3}^T} {\Delta m_D} {\widetilde
  N_{7\times 3}} + {\widetilde N_{3\times 3}^T} {\Delta m} {\widetilde
  N_{3\times 3}}.
\label{flavour-mass-relation}
\eeq
Typically, for a Dirac neutrino, the mass is $\lsim$ $\cal{O}$
$(10^{-4}~\rm{GeV})$, while for a neutralino, the mass is 
$\sim$ $\cal{O}$ $(10^{2}~\rm{GeV})$. This means that the 
entries of the off-diagonal blocks in
eq.(\ref{neutralino-mixing-matrix-block-form}) are $\lsim$ $\cal{O}$
$(10^{-6})$. Therefore, for all practical purposes, one can neglect
the first three terms in comparison to the fourth term on the right
hand side of eq.(\ref{flavour-mass-relation}). Thus,
\beq
\Delta m_{f} \approx {\widetilde N_{3\times 3}^T} {\Delta m} {\widetilde N_{3\times 3}}.
\label{flavour-mass-relation-2}
\eeq
up to a very good approximation. With this in view, our strategy is to compute
the one-loop corrections in the $\widetilde \chi^0$ basis first, and then use
eq.(\ref{flavour-mass-relation-2}) to obtain the corresponding corrections in
the flavour basis. Finally, we diagonalize
eq.(\ref{mass-basis-seesaw-schematic-5}) to obtain the one-loop corrected
neutrino masses. We have performed all calculations in the 't-Hooft-Feynman
gauge. Let us also note in passing that the form of eq.(\ref{one-loop-corrected-mass}) 
predicts off-diagonal entries ($i\neq j$). The off-diagonal elements are
 responsible for the mixing between diagonal entries, they become dominant
 only when $\left({m}_{\n_i}-{m}_{\n_j}\right)~\lesssim(\frac{\alpha}{4 \pi})
 \times {\rm{some ~electroweak ~scale ~mass}}$, and then, one can choose $p^2 =
 \ovl{m^2} = ({{m^2}_{\n_i} + {m^2}_{\n_j}})/2$ for external momentum\cite{Hempfling-Rpv}. Thus,
 one can conclude that unless the tree level masses are degenerate, the
 off-diagonal radiative corrections can be neglected for all practical
 purposes, when at least one index $i$ or $j$ refers to a heavy states.

The self-energy corrections contain entries of the neutralino mixing matrix
$N$ through the couplings $O^{ff^{\prime}b}$ (see, appendix E). This is because, the 
self energies ${\it{\widetilde \Sigma}_{ij}}$ and ${\it{\widetilde \Pi}_{ij}}$ in
general contain products of couplings of the form
$O^{ff^{\prime}b}_{i..}O^{ff^{\prime}b}_{j..}$. The matrix $N$, on the other
hand, contains the expansion parameter $\xi$ in the leading order (see
eq.(\ref{neutralino-mixing-matrix})). This observation, together with the help
of eq.(\ref{expansion-parameter-terms}), help us to express the
effective structure of the one-loop corrected neutrino mass matrix as,
\beq
[(\mathcal{M}^{\nu^{\prime}})_{eff}]_{ij} = A_1 a_i a_j + A_2 c_i c_j + A_3 (a_i c_j + a_j c_i),
\label{one-loop corrected structure of neutralino mass matrix}
\eeq
where $a_i$ and $c_i$ are given by eq.(\ref{specifications}) and
$A_i$'s are functions of our model parameters and the Passarino-Veltman
functions $(B_0,B_1)$ \cite{Passarino-Veltman,Veltman-tHooft,Hahn-Victoria} defined in 
appendix \ref{The $B_0$ and $B_1$ function}. The form of the loop corrected mass matrix thus 
obtained is identical to the tree level one (see, eq.(\ref{mnuij-compact-recasted}))
with different coefficients $A_1$, $A_2$ and $A_3$ arising now.  

Note that the one-loop diagrams in figure\ref{one-loop-diagrams}, contributing to the neutrino mass matrix 
are very similar to those obtained in bilinear R-parity violating scenario \cite{Hirsch-loop-FD,Valle-Porod-others,
Davidson-Losada-jhep,Davidson-Losada-prd,Grossman-Rakshit-prd,Dedes-Rimmer-Rosiek}. However, it has been pointed
out in Ref.\cite{Porod-Bartl}, that there is a new significant contribution coming from the loops containing the
neutral scalar and pseudoscalar with dominant singlet component. This contribution is proportional to the 
mass-splitting between the singlet scalar and pseudoscalar states \cite{Hirsch-others,grossman-haber-prl,Dedes-Haber-Rosiek}.  
The corresponding mass splittings for the doublet sneutrinos are much smaller \cite{Porod-Bartl}.

\FIGURE{\epsfig{file=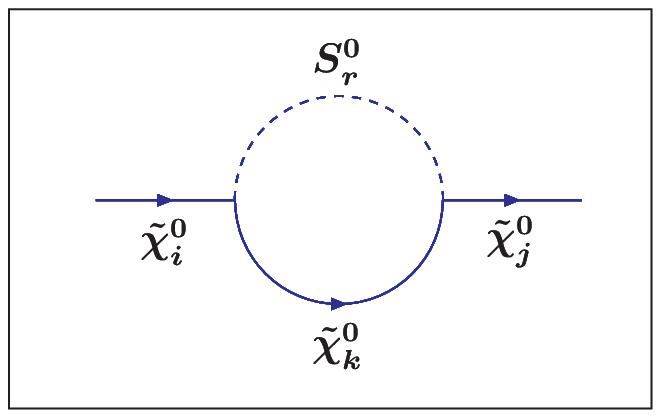,height=2.75cm} 
\vspace{0.2cm}
\epsfig{file=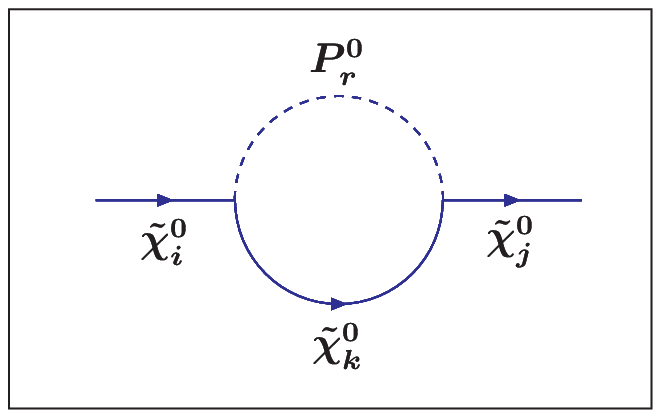,height=2.75cm} 
\vspace{0.2cm}
\epsfig{file=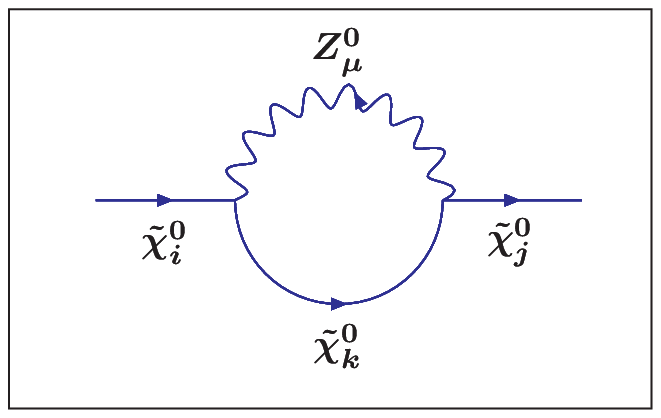,height=2.75cm} 
\vspace{0.2cm}
\epsfig{file=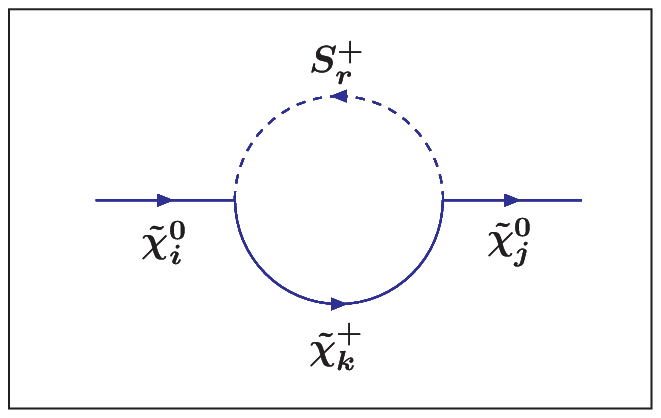,height=2.75cm} 
\vspace{0.2cm}
\epsfig{file=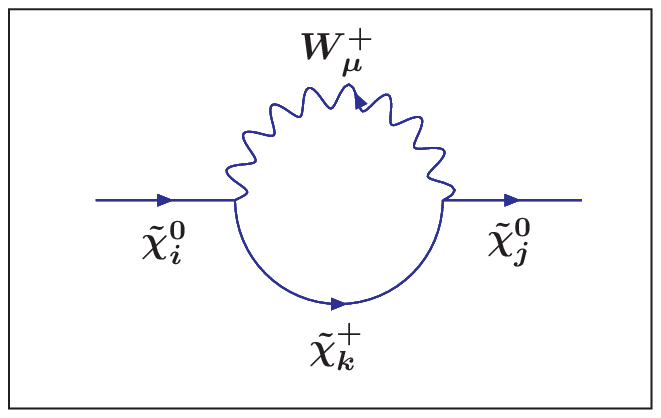,height=2.75cm} 
\vspace{0.2cm}
\epsfig{file=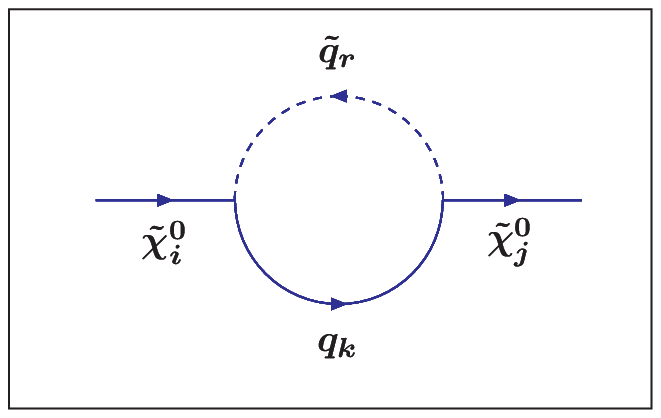,height=2.75cm}
\caption{One-loop diagrams contributing to the neutralino masses. The various
  contributions are arising from (clockwise from top left) 
  (a) neutralino-neutralino-neutral scalar loop, 
  (b) neutralino-neutralino-neutral pseudoscalar loop, 
  (c) neutralino-neutralino-$Z^0_{\mu}$ loop, 
  (d) neutralino-chargino-charged scalar loop, 
  (e) neutralino-chargino-$W^{\pm}_{\mu}$ loop, 
  (f) neutralino-quark-squark loop.}
\label{one-loop-diagrams}}


\subsection{Neutrino mixing}
The unitary matrix which diagonalizes the $3\times3$ light neutrino
mass matrix, can be parametrized as \cite{neutrino_pmns}, 

\bea
{\mathcal{U}_\nu} &=&
\left(\begin{array}{ccc}
{c_{12}}{c_{13}} & {s_{12}}{c_{13}} & {s_{13}} \\ \\
-{s_{12}}{c_{23}}-{c_{12}}{s_{23}}{s_{13}} & {c_{12}}{c_{23}}
-{s_{12}}{s_{23}}{s_{13}}  & {s_{23}}{c_{13}}\\ \\
{s_{12}}{s_{23}}-{c_{12}}{c_{23}}{s_{13}} & -{c_{12}}{s_{23}}
-{s_{12}}{c_{23}}{s_{13}}  & {c_{23}}{c_{13}}
\end{array}\right),
\label{neutrino_mixing_3x3}
\eea 
provided the charged lepton sector is in the mass-basis (see appendix
\ref{Charged fermion mass matrix}). Here $c_{ij} = \cos{\theta_{ij}}$,
$s_{ij} = \sin{\theta_{ij}}$, and all $CP$ violating phases (Dirac or
Majorana) are set to zero. The experimental data on neutrino
oscillations \cite{osc_experiments1, osc_experiments2} indicate that a
bilarge pattern of mixing of neutrinos may be preferred, which means
that two of the mixing angles must be large. As a first approximation,
one can work with $\theta_{23} = 45^\circ$, $\sin{\theta_{12}} =
{\frac{1}{\sqrt{3}}}$ and $\theta_{13} \approx 0^\circ$, which is
often referred to as the `tribimaximal structure'
\cite{tribimaximal}. 
Following the above discussions, we can write down the tree level PMNS
matrix ${\mathcal{U}_{\nu}}$ as,
\beq\label{U_pmns-tree}
{\mathcal{U}_{\nu}} = {\widetilde N^{{\prime}^T}}_{3\times3},
\eeq
where ${\widetilde N^{{\prime}}}_{3\times3}$ is the matrix that diagonalizes
the tree level mass matrix $M^{seesaw}_{\nu}$
(eq.(\ref{seesaw_formula})).

Similarly, the $3\times3$ unitary matrix that
diagonalizes the one-loop corrected neutrino mass matrix
$({M}^{\nu^{\prime}})_{\rm{eff}}$
(eq.(\ref{mass-basis-seesaw-schematic-5})), can be denoted as
$\mathcal{U}^{\prime}_\nu$. 
Symbolically
\beq
\label{diag-one-loop-corrected-neutrino-mass}
{\mathcal{U}^{{\prime}^{-1}}_{\nu}} ({M}^{\nu^{\prime}})_{\rm{eff}}
{\mathcal{U}^{\prime}_{\nu}} ={\rm{diag}} (m'_1,~m'_2,~m'_3),
\eeq
with $m'_1$, $m'_2$, $m'_3$ as the three one-loop corrected light neutrino
masses. A similar relation for the tree level calculation is given by the
second equation of (\ref{diag-matrix}), with $m_i$'s as the tree level neutrino
masses. 

When we include one-loop corrections,
the PMNS matrix ${\mathcal{U}^{\prime}_{\nu}}$ is defined as 
\beq
\label{U_pmns-one-loop}
{\mathcal{U}^{\prime}_{\nu}} = {\widetilde N^{{\prime\prime}^T}}_{3\times3},
\eeq
where ${\widetilde N^{{\prime\prime}}}_{3\times3}$ is the matrix which
diagonalizes the one-loop corrected mass matrix $({M}^{\nu^{\prime}})_{\rm{eff}}$. 
The scheme for obtaining the loop-corrected neutrino mixing matrix is shown
in figure\ref{flowchart}.

\FIGURE{\epsfig{file=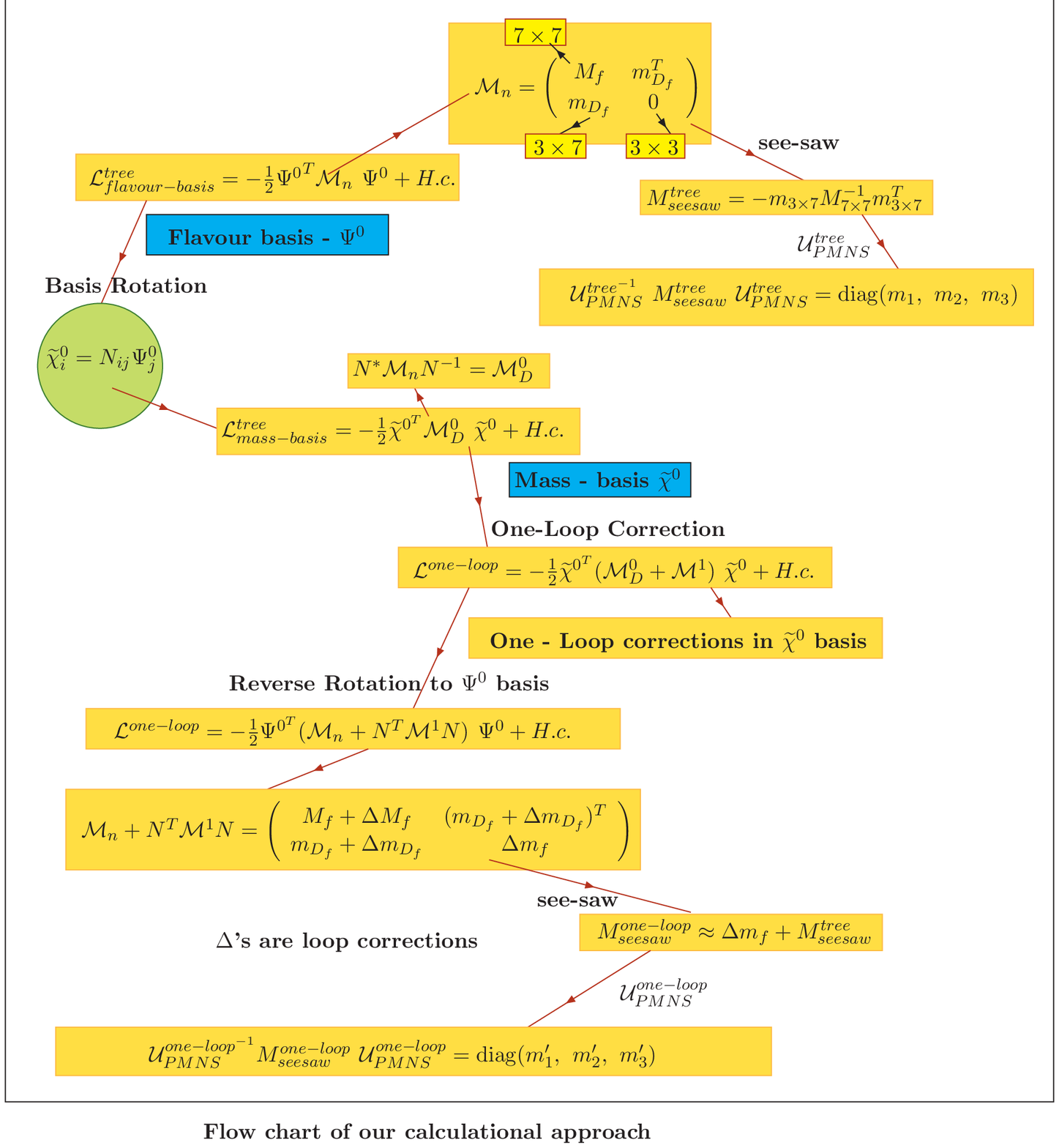,height=16.55cm} 
\caption{Flow chart of the calculational procedure that we have followed in order
to calculate the neutrino masses and mixing at the tree level and at one-loop. Here 
${\mathcal{U}^{one-loop}_{PMNS}}$  = ${\mathcal{U}^{\prime}_{\nu}}$, ${\mathcal{U}^{tree}_{PMNS}}$ 
= ${\mathcal{U}_{\nu}}$, $\mathcal{L}^{tree}_{flavour-basis} = \mathcal{L}^{mass}_{neutral}$, 
$M^{tree}_{seesaw} = M^{seesaw}_{\nu}$, $\mathcal{L}^{one-loop} = \mathcal{L}^{\prime}$ and  
$M^{one-loop}_{seesaw} = (M^{\nu^{\prime}})_{eff}$ in the text.}
\label{flowchart}}
\section{Numerical results of neutrino mass and mixing}
\label{Numerical results of neutrino mass and mixing}
In this section we present and explain the results of our numerical
analysis. Let us begin with a brief outline of the different well known
schemes of light neutrino masses as favoured by experiments. These are (i) The
normal hierarchy: $m_1 < m_2 \sim \sqrt{\Delta m^2_{solar}}$ ,~$m_3 \sim
\sqrt{|\Delta m^2_{atm}|}$, (ii) The inverted hierarchy: $m_1 \approx m_2 \sim
\sqrt{|\Delta m^2_{atm}|}$, $m_3 \ll \sqrt{|\Delta m^2_{atm}|}$ and (iii) The
quasi-degenerate pattern: $m_1 \approx m_2 \approx m_3 \gg \sqrt{|\Delta
  m^2_{atm}|}$, where $m_1$, $m_2$ and $m_3$ are the three light neutrino
masses. Here, $\Delta m^2_{solar} \equiv m^2_2 - m^2_1$ and 
~$\Delta m^2_{atm} \equiv m^2_3 - m^2_2 (m^2_1)$ for Normal (inverted) Hierarchy.
The possibility of more than one scheme of neutrino
masses essentially stems from our lack of knowledge of the signs of the
squared mass differences, or the value of individual masses. Such being the
case, the numerical analysis must be subjected to address all the three
probable schemes.

The numerical calculations have been performed with the help of a code
developed by us in Fortran and the results have been cross-checked using 
another Mathematica \cite{math-wolfram} based program we developed.
In the code, we keep the left handed sneutrino VEVs ($v'$), the right handed
sneutrino VEVs $(v^c)$ and the neutrino Yukawa couplings $(Y_\nu)$ as free
input variables, and then scan over the parameter space for a region
consistent with the three flavour global neutrino data. For all our numerical
analysis we keep the right handed sneutrino VEVs fixed at some chosen values
(see table \ref{table1}). The only exception to this is when we study the
correlation of neutrino data with the bilinear $R_P$-violating parameter
$\varepsilon_i=Y^{ij}_\nu v^c_j$. For that particular study, we vary the
right handed sneutrino VEVs (consistent with EWSB conditions) within 
the mass scale --$895 ~\rm{GeV}$ to --$565 ~\rm{GeV}$. 
The table \ref{table1} shows our choice of the sample parameters for the numerical
analysis. The relation between the gaugino soft masses $M_1$ and $M_2$ are
assumed to be GUT (grand unified theory) motivated, so that, at the
electroweak scale, we have $M_1 : M_2~=~1:2$. We choose $M_1~=~110~\rm{GeV}$.

\TABLE {\begin{tabular}{|c|c|c|c|c|c|c|c|} \hline \rm{tan}$\beta$ & $\lambda$
    & $A_\lambda \lambda$ & $\kappa$ & $A_\kappa \kappa$ & $(A_\nu
    Y_\nu)^{ii}$ & $(m_{\tilde e^c}^2)$ & $v^c_i$\\ \hline 10 & 0.10 &
    $-1\rm{TeV} \times \lambda$ & 0.45 & $1\rm{TeV} \times \kappa$ &
    $1\rm{TeV} \times Y^{ii}_\nu$ & $300^2 ~\rm{GeV}^2$ & $-595~\rm{GeV}$\\ \hline
\end{tabular}
\caption{Choice of parameters for numerical analysis consistent with the EWSB
  conditions.\label{table1}}}

We scanned the parameter space comprising of the left handed sneutrino VEVs and 
the neutrino Yukawa couplings extensively and found certain ranges of these parameters 
appropriate for various hierarchical schemes of the light neutrino masses. 
In table \ref{table2}, some sample values for these six 
parameters for the different mass schemes of neutrinos are
given. These values are just for illustration and the parameters were
scanned around these numbers to generate the plots shown in this
section. The other relevant parameters have values as mentioned in
table \ref{table1}.

\TABLE[t]
{\begin{tabular}{|c|c|c|c|c|c|c|}
\hline
 & \multicolumn{3}{c|}{$Y^{ii}_\nu \times 10^7$} 
& \multicolumn{3}{c|}{$v'_i \times 10^5 (\rm{GeV})$ } \\
\cline{2-7}
 & $Y^{11}_\nu$ & $Y^{22}_\nu$ & $Y^{33}_\nu$ & $v'_1$ 
& $v'_2$ & $v'_3$ \\
\hline
\text{Normal hierarchy} & 3.550 & 5.400 & 1.650 & 0.730 & 10.100 & 12.450\\
\hline
\text{Inverted hierarchy} & 12.800  & 3.300 & 4.450 & 8.350  & 8.680 & 6.400\\
\hline
\end{tabular}
\caption{Values of neutrino Yukawa couplings and left handed sneutrino VEVs,
  used as sample parameter points for numerical calculations. These are the
  values around which the corresponding parameters were
  varied. \label{table2}}}

While fitting the three flavour global neutrino data, we consider constraints 
arising from the oscillation data as well as from the non-oscillation data. 
We probe the effects of these constraints for both tree level and (tree + one-loop) 
level analyses. The oscillation data constrain the solar and atmospheric mass 
squared differences, namely, $\Delta m^2_{solar}$ and $\Delta m^2_{atm}$, and 
three neutrino mixing angles $\theta_{13},~\theta_{12},~\theta_{23}$. The present 
3$\sigma$ limits are \cite{boris_review_08, Schwetz-Valle},
\bea
7.05 \times 10^{-5} {\rm eV}^2 \leq \Delta m^2_{solar} & &\leq 8.34 \times 10^{-5} 
{\rm eV}^2, \nonumber \\
2.07 \times 10^{-3} {\rm eV}^2 \leq |\Delta m^2_{atm}| & &\leq 2.75 \times 
10^{-3} {\rm eV}^2,\nonumber\\
0.25 \leq \sin^2\theta_{12} & &\leq 0.37, \nonumber\\
0.36 \leq \sin^2\theta_{23} & &\leq 0.67, \nonumber\\
\sin^2\theta_{13} & &\leq 0.056. \nonumber\\ 
\label{oscillation-data}
\eea
The non-oscillation constraints follow from experiments like 
$\beta$ decay \cite{Bonn:2001tw,Lobashev:2001uu,Osipowicz:2001sq},
neutrinoless double beta decay 
$(0\nu\beta\beta)$ \cite{KlapdorKleingrothaus:2004wj,KlapdorKleingrothaus:2006ff,Arnaboldi:2008ds}
(this is also sensitive to Majorana nature and phases), and from 
cosmology \cite{Komatsu:2008hk}. Here we set all the Majorana phases to 
be zero, as we are dealing with a $CP-$preserving situation.

The set of non-oscillation constraints are given as,
\bea 
m_\beta = \sqrt{\sum|{{\mathcal{U}_{\nu_{ei}}}}|^2 m^2_i} & &< 1.80
~\rm{eV} ~~~~(\beta ~decay), \nonumber \\ 0.00 ~{\rm{eV}}\leq m_{\beta\beta} =
|\sum{{\mathcal{U}_{\nu_{ei}}}}^2 m_i| & &\leq 0.25 ~\rm{eV}
~~~~(0\nu\beta\beta),\nonumber\\ \sum m_i & &< 1.30 ~\rm{eV} ~~~~(cosmology),
\nonumber\\
\label{non-oscillation-data}
\eea
where $m_i$s are three light neutrino masses, and ${\mathcal{U}_{\nu_{ei}}}$s
are the elements of the first row of the neutrino mixing matrix (see
eq.(\ref{neutrino_mixing_3x3})).

\subsection{Normal hierarchy}
\label{Normal hierarchy}
In the normal hierarchical pattern of the three light neutrino masses, 
the atmospheric and the solar mass squared differences, given by 
$\Delta m^2_{atm}= m^2_3 - m^2_2$ and $\Delta m^2_{solar}= m^2_2 - m^2_1$, 
are largely governed by the higher mass squared in each case, namely, 
$m_3^2$ and $m_2^2$, respectively. Before going into the discussion of 
the variation of the mass-squared values with the model parameter, some 
general remarks are in order. First of all, note that in eq.(\ref{specifications}), 
if we choose $v^\prime_i$ such that $v'_i \gg \frac{Y^{ii}_\nu v_1}{3 \lambda}$, 
then $b_i \approx c_i$\cite{munoz-lopez-3}. Thus, for large $v^\prime_i$,
we remind ourselves that in eq.(\ref{mnuij-compact1}), the effective light neutrino
mass matrix has two types of seesaw structures \cite{Ghosh-Roy}. The first 
one is the {\it{ordinary seesaw}}, given by
\bea 
m_\nu \sim \frac{a^2_{i}}{m_{\nu^c}},
\eea 
where $a_i = Y_{\nu}^{ii} v_2$ represents the {\it{Dirac}} mass term 
for neutrinos, and $m_{\nu^c} = {2 \kappa v^c}$ stands for the
{\it{Majorana}} mass term of the right handed neutrino. The second
type is called the {\it{gaugino seesaw}}, in which the role of the
{\it{Dirac}} mass terms are played by $g_1 c_i$ and $g_2 c_i$, where
$g_1,~g_2$ are the $U(1)$ and the $SU(2)$ gauge couplings respectively
and $c_i$ stands for the left handed sneutrino VEV $v'_i$. The
role of the {\it{Majorana}} masses are played by the gaugino soft
masses $M_1,~M_2$. This seesaw relation is given as 
\bea
m_\nu &\sim& \frac{(g_1
  c_i)^2}{M_1} + \frac{(g_2 c_i)^2}{M_2}, \nonumber \\ 
 &\sim&\frac{c^2_i}{M},
\eea
where the subscript `$i$' $=1,2,3 \equiv e, \mu, \tau$ and $M$ is
the reduced gaugino mass defined by
\bea
\frac{1}{M} = \frac{g^2_1}{M_1} + \frac{g^2_2}{M_2}.
\eea 
$M$ here plays the role of the effective heavy mass provided by
the neutral electroweak gaugino sector, and the effect is closely
analogous to Type-III seesaw mechanism\cite{type-3-seesaw}\footnote
{We thank Anjan Joshipura for pointing this out to one of the authors in a 
private discussion.}. As discussed after 
eq.(\ref{one-loop corrected structure of neutralino mass matrix}),
when one-loop corrections are added, the neutrino masses are still 
determined by the quantities $a^2_i$ and $c^2_i$ ($\approx b^2_i$ 
for large $v^\prime_i$).

\FIGURE{\epsfig{file=Figures1/masssqesgs1s.eps,height=4.55cm} 
\vspace{0.2cm}
\epsfig{file=Figures1/masssqesos1s.eps,height=4.55cm} 
\vspace{0.2cm}
\epsfig{file=Figures1/masssqesgs2s.eps,height=4.55cm} 
\vspace{0.2cm}
\epsfig{file=Figures1/masssqesos2s.eps,height=4.55cm} 
\vspace{0.2cm}
\epsfig{file=Figures1/masssqesgs3s.eps,height=4.55cm} 
\vspace{0.2cm}
\epsfig{file=Figures1/masssqesos3s.eps,height=4.55cm}
\caption{Neutrino mass squared values ($m^2_i$) vs $\frac{c^4_i}{M^2}$ 
(left panel) and vs $\frac{a^4_i}{m^2_{\nu^c}}$ (right panel) plots for 
the {\it{normal hierarchical}} pattern of light neutrino masses, $i = e,
 \mu, \tau$.}
\label{numsqNH}}

In the subsequent plots, we show the variation of the neutrino squared 
masses ($m^2_i$) and the atmospheric and solar mass squared differences 
with the square of the seesaw parameters $\frac{c^2_i}{M}$ and 
$\frac{a^2_{i}}{m_{\nu^c}}$. Results are shown for the tree level 
as well as the one-loop corrected neutrino masses. These plots also 
demonstrate the importance of one-loop corrections to neutrino masses 
compared to the tree level results.  

Typical mass spectra are shown in figure \ref{numsqNH}. Note that a 
particular model parameter has been varied while the others are fixed
at values given in tables \ref{table1} and \ref{table2}. The effective 
light neutrino mass matrix given in eq.(\ref{mnuij-compact1}) 
suggests that as long as $v'_i \gg \frac{Y^{ii}_\nu v_1}{3 \lambda}$ and 
$\kappa \gg \lambda$, the second term on the right hand side of 
eq.(\ref{mnuij-compact1}) dominates over the first term and as a result 
the heaviest neutrino mass scale ($m_3$) is controlled mainly by the gaugino 
seesaw effect. This is because in this limit $b_i \approx c_i$, and, as discussed 
earlier, a neutrino mass matrix with a structure $(m_\nu)_{ij} \sim \frac{c_ic_j}{M}$ 
can produce only one non-zero neutrino mass. This feature is evident in 
figure\ref{numsqNH}, where we see that $m^2_3$ increases as a function of 
$c^4_i/M^2$. The other two masses are almost insensitive to $c^2_i/M$. A mild variation 
to $m^2_2$ comes from the combined effect of gaugino and ordinary seesaw.
On the other hand, the two lighter neutrino mass scales ($m^2_2$ and
$m^2_1$) are controlled predominantly by the ordinary seesaw parameters
$a^2_i/{m_{\nu^c}}$. This behaviour is observed in the right panel figures of
figure\ref{numsqNH}. The heaviest neutrino mass scale is not much affected
by the quantities $a^2_i/{m_{\nu^c}}$.    

One can also see from these plots that the inclusion of one-loop 
corrections, for the chosen values of the soft SUSY breaking parameters,
reduces the values of $m^2_2$ and $m^2_1$, while increasing the 
value of $m^2_3$ only mildly. This is because, with such a choice,  the 
one-loop corrections cause partial cancellation in the generation of $m_1$ 
and $m_2$. For the heaviest state, it is just the opposite, since the 
diagonalization of the tree-level mass matrix already yields 
a negative mass eigenvalue, on which the loop correction has an additive effect. 
If, with all other parameters fixed, the signs of $\lambda$ and $A_\lambda$ are 
reversed (leading to a positive $\mu$ in the place of a negative one), $m_1$, 
$m_2$ and $m_3$ are all found to decrease through loop corrections. A flip in the
sign of $\kappa$ and the corresponding soft breaking terms, on the other hand, causes
a rise in all the mass eigenvalues, notably for $m_1$ and $m_2$.
 
In the light of the discussion above, we now turn to explain the 
variation of $\Delta m^2_{atm}$ and $\Delta m^2_{solar}$
with $c^4_i/M^2$ and $a^4_i/{m^2_{\nu^c}}$ shown in figure\ref{gsNH} 
and figure\ref{osNH}. For our numerical analysis, in order to 
set the scale of the normal hierarchical spectrum, we choose 
$m_2|_{max}<0.011~\rm{eV}$. The left panel in 
figure\ref{gsNH} shows that $\Delta m^2_{atm}$ increases more 
rapidly with $c^4_{\mu,\tau}/M^2$, whereas the variation with 
$c^4_e/M^2$ is much slower as expected from figure\ref{numsqNH}. 
Similar behaviour is shown for the one-loop corrected 
$\Delta m^2_{atm}$. The small increase in the one-loop 
corrected result compared to the tree level one is essentially due 
to the splitting in $m^2_2$ value as shown earlier. The variation 
of $\Delta m^2_{solar}$ with $c^4_i/M^2$ 
can be explained in a similar manner. Obviously, in this case
the one-loop corrected result is smaller compared to the tree 
level one (see, figure\ref{numsqNH}). However, one should note
that $\Delta m^2_{solar}$ falls off with $c^4_{\mu}/M^2$ as 
opposed to the variation with respect to the other two 
gaugino seesaw parameters. This is due to the fact that $m^2_2$ 
slightly decreases with $c^4_{\mu}/M^2$ but show a slow increase 
with respect to $c^4_{e}/M^2$ and $c^4_{\tau}/M^2$. The dark solid 
lines in all these figures show the allowed values of various parameters
where all the neutrino mass and mixing constraints are satisfied.

\FIGURE{\epsfig{file=Figures1/atmsqvsgs1sNH.eps,height=4.55cm} 
\vspace{0.2cm}
\epsfig{file=Figures1/solsqvsgs1sNH.eps,height=4.55cm} 
\vspace{0.2cm}
\epsfig{file=Figures1/atmsqvsgs2sNH.eps,height=4.55cm} 
\vspace{0.2cm}
\epsfig{file=Figures1/solsqvsgs2sNH.eps,height=4.55cm} 
\vspace{0.2cm}
\epsfig{file=Figures1/atmsqvsgs3sNH.eps,height=4.55cm} 
\vspace{0.2cm}
\epsfig{file=Figures1/solsqvsgs3sNH.eps,height=4.55cm}
\caption{Atmospheric and solar mass squared differences $(\Delta
  m^2_{atm},~\Delta m^2_{solar})$ vs $\frac{c^4_i}{M^2}$ plots for the
  {\it{normal hierarchical}} pattern of light neutrino masses, $i = e,
  \mu, \tau$. The full lines are shown for which only the constraints on 
  $\Delta m^2_{solar}$ is not within the 3$\sigma$ limit. The dark 
  coloured portions on these lines are the values of parameters 
  for which all the neutrino constraints are within the 3$\sigma$ limit. The 
  red (yellow) coloured lines in the plots correspond to the tree (one-loop corrected) 
  regions where all the constraints except $\Delta m^2_{solar}$ are within 3$\sigma$ 
  allowed region. Other parameter choices are discussed in the text.}
\label{gsNH}}

\FIGURE{\epsfig{file=Figures1/atmsqvsos1sNH.eps,height=4.55cm} 
\vspace{0.2cm}
\epsfig{file=Figures1/solsqvsos1sNH.eps,height=4.55cm} 
\vspace{0.2cm}
\epsfig{file=Figures1/atmsqvsos2sNH.eps,height=4.55cm} 
\vspace{0.2cm}
\epsfig{file=Figures1/solsqvsos2sNH.eps,height=4.55cm} 
\vspace{0.2cm}
\epsfig{file=Figures1/atmsqvsos3sNH.eps,height=4.55cm} 
\vspace{0.2cm}
\epsfig{file=Figures1/solsqvsos3sNH.eps,height=4.55cm}
\caption{Atmospheric and solar mass squared differences $(\Delta
  m^2_{atm},~\Delta m^2_{solar})$ vs $a^4_i/m^2_{\nu^c}$ plots for the
  {\it{normal hierarchical}} pattern of light neutrino masses with $i
  = e, \mu, \tau$. Colour specification is same as described 
  in the context of figure \ref{gsNH}. Other parameter choices are discussed 
  in the text.}
\label{osNH}}

The variation of $\Delta m^2_{atm}$ and $\Delta m^2_{solar}$ with 
$a^4_i/m^2_{\nu^c}$ in figure\ref{osNH} can be understood in a 
similar way by looking at the right panel plots of 
figure\ref{numsqNH}. $\Delta m^2_{atm}$ shows a very little increase 
with $a^4_{e,\mu}/m^2_{\nu^c}$ as expected, whereas the change is more 
rapid with $a^4_\tau/m^2_{\nu^c}$ for the range of values considered
along the x-axis. As in the case of figure\ref{gsNH}, the solid dark lines
correspond to the allowed values of parameters where all the neutrino 
mass and mixing constraints are satisfied. 

For higher values of $a^4_{e,\tau}/m^2_{\nu^c}$, $m^2_2$ increases 
very slowly with these parameters (see, figure\ref{numsqNH}) and this is
reflected in the right panel plots of figure\ref{osNH}, where 
$\Delta m^2_{solar}$ shows a very slow variation with 
$a^4_{e,\tau}/m^2_{\nu^c}$. On the other hand, $m^2_2$ increases 
more rapidly with $a^4_\mu/m^2_{\nu^c}$, giving rise to a faster
variation of $\Delta m^2_{solar}$. The plots of figure\ref{osNH} show 
that larger values of Yukawa couplings are required in order to satisfy
the global three flavour neutrino data, when one considers one-loop
corrected neutrino mass matrix. However, there are allowed ranges of
the parameters $a^4_i/m^2_{\nu^c}$, where the neutrino data can be 
satisfied with both tree and one-loop corrected analysis.  

We have also considered the variation of light neutrino mass 
squared differences with the effective bilinear $R_P$
violating parameter, $\varepsilon_i = Y^{ij} v^c_j$. 
For this particular numerical study we vary both
$Y^{ii}_\nu$ and the right handed sneutrino VEVs $v^c_i$ 
simultaneously, in the suitable ranges around the
values given in table \ref{table1} and \ref{table2}. 
$\Delta m^2_{atm}$ is found to increase
with $\varepsilon_i$, whereas the solar mass squared
difference decreases with increasing $\varepsilon_i$. 
The $3\sigma$ allowed region for the solar and atmospheric mass 
squared differences were obtained for the lower values of 
$\varepsilon_i$s. In addition, we have noticed that the correlations 
of $\Delta m^2_{atm}$ with $\varepsilon_i$ is sharper compared to 
the correlations seen in the case of $\Delta m^2_{solar}$.

Next let us discuss the dependence of $\Delta m^2_{atm}$ and 
$\Delta m^2_{solar}$ on two specific model parameters, $\lambda$ and 
$\kappa$, consistent with EWSB conditions. The loop corrections
shift the allowed ranges of $\kappa$ to lower values with some 
amount of overlap with the tree level result. On the other hand,  
the allowed ranges of $\lambda$ shrinks towards higher values when
one-loop corrections are included. These results are shown in 
figure\ref{L-K-NH}. We note in passing that the mass of the  
lightest CP-even scalar decreases with increasing $\lambda$. For example, 
$\lambda = 0.15$ can produce a lightest scalar mass of $40 ~\rm{GeV}$, 
for suitable choices of other parameters. This happens
because with increasing $\lambda$, the lightest scalar state picks up
more and more right handed sneutrino admixture. 

\FIGURE{\epsfig{file=Figures1/atmosvskappaNH.eps,height=4.55cm} 
\vspace{0.2cm}
\epsfig{file=Figures1/solarvskappaNH.eps,height=4.55cm} 
\hspace{8.4cm}
\epsfig{file=Figures1/atmosvslambdaNH.eps,height=4.55cm} 
\vspace{0.2cm}
\epsfig{file=Figures1/solarvslambdaNH.eps,height=4.55cm} 
\caption{Plots showing the variations of $\Delta m^2_{atm},~\Delta
  m^2_{solar}$ with model parameters $\lambda$ and $\kappa$ for
  {\it{normal hierarchy}}. Values of all other parameters are given by
  table \ref{table1} and \ref{table2}. Colour specification is same as 
  described in the context of figure \ref{gsNH}.
  Other parameter choices are discussed in the text.}
\label{L-K-NH}}

Finally, we will discuss the $\rm{tan}\beta$ dependence of $\Delta
m^2_{atm}$ and $\Delta m^2_{solar}$. These plots are shown in figure 
\ref{tanbetaNH}. The quantity $\Delta m^2_{atm}$ decreases with the
increasing values of $\rm{tan}\beta$ and nearly saturates for larger 
values of $\rm{tan}\beta$. However, the one-loop corrected result 
for $\Delta m^2_{atm}$ is not much different from that at the 
tree level for a particular value of $\tan\beta$. 
On the other hand, the solar mass squared difference initially 
increases with $\tan\beta$ and for higher values of $\tan\beta$ 
the variation slows down and tends to saturate. The one-loop corrections 
result in lower values of $\Delta m^2_{solar}$ for a particular $\tan\beta$. 
The darker and bigger points on both the plots of figure\ref{tanbetaNH} are 
the allowed values of $\tan\beta$, where all the neutrino experimental data 
are satisfied. Note that only a very small range of $\tan\beta$ ($\sim$ 10--14) 
is allowed. This is a very important observation of this analysis. 

\FIGURE{\epsfig{file=Figures1/atmsqvstanbNH.eps,height=4.55cm} 
\vspace{0.2cm}
\epsfig{file=Figures1/solsqvstanbNH.eps,height=4.55cm} 
\caption{$\Delta m^2_{atm},~\Delta m^2_{solar}$ vs $\rm{tan}\beta$
  plots for the {\it{normal hierarchical}} pattern of light neutrino
  masses. The allowed values of $\tan\beta$ are shown by bold points. 
  Other parameter choices are shown in table \ref{table1} 
  and \ref{table2}.}
\label{tanbetaNH}}

Next we will discuss the light neutrino mixing and the effect of 
one-loop corrections on the mixing angles. It was shown in 
ref.\cite{Ghosh-Roy} that for the normal hierarchical pattern of 
neutrino masses, when the parameter $b_i \sim a_i$, the 
neutrino mixing angles $\theta_{23}$ and $\theta_{13}$ can be 
written as (with the tree level analysis),
\bea
\sin^2\theta_{23} \approx \frac{b^2_\mu}{b^2_\mu + b^2_\tau},
\eea
and 
\bea
\sin^2\theta_{13} \approx \frac{b^2_e}{b^2_\mu + b^2_\tau}.
\eea
On the other hand, the mixing angle $\theta_{12}$ is a much more 
complicated function of the parameters $b_i$ and $a_i$ and we do 
not show it here. Now, when $b_i \sim a_i$, we can easily 
see from eq.(\ref{specifications}), that
\bea
v^\prime_i \sim \frac{Y^{ii}_\nu v_1}{3\lambda}(\tan\beta -1).
\eea 
This implies that for $\tan\beta \gg$ 1 (recall that the allowed 
range of $\tan\beta$ is $\sim$ 10--14), 
\bea
v^\prime_i \gg \frac{Y^{ii}_\nu v_1}{3\lambda}.
\eea 
As we have discussed earlier, for such values of $v^\prime_i$, 
the quantities $b_i \approx c_i$. Hence, the mixing angles 
$\theta_{23}$ and $\theta_{13}$ can be approximately written as
\bea
\sin^2\theta_{23} \approx \frac{c^2_\mu}{c^2_\mu + c^2_\tau},
\label{eq-mixing-23} 
\eea
and
\bea
\sin^2\theta_{13} \approx \frac{c^2_e}{c^2_\mu + c^2_\tau}.
\label{eq-mixing-13} 
\eea
Naively, one would also expect that $\sin^2\theta_{12}$ should
show some correlation with the quantity $c^2_e/c^2_\mu$. However,
as mentioned earlier, this is a very simple minded expectation since
$\sin^2\theta_{12}$ has a more complicated dependence on the model
parameters. 

The variation of all three mixing angles with the corresponding 
parameters are shown in figure\ref{mixing-GS-OS}. Note that in order 
to generate these plots, we vary only the quantities $c_i$ and 
all the other parameters are fixed at the values given in 
tables \ref{table1} and \ref{table2}. We have chosen the
range of parameters in such a way that the 3-flavour global neutrino 
data are satisfied. The mixing angles have been calculated numerically 
by diagonalizing the neutrino mass matrix in eq.(\ref{mnuij-compact1}) 
and in eq.(\ref{one-loop corrected structure of neutralino mass matrix}). 
As expected from our approximate analytical expressions, these plots 
show very nice correlations of the mixing angles $\theta_{23}$ and 
$\theta_{13}$ with the relevant parameters as discussed in 
eqs.(\ref{eq-mixing-23}) and (\ref{eq-mixing-13}). For example, note 
that when $c_\mu \approx c_\tau$, $\sin^2\theta_{23}$ is predicted to 
be $\approx$ 0.5 and that is what we observe in the tree level plot in 
figure\ref{mixing-GS-OS}. However, when one-loop corrections are 
considered, the value of $\sin^2\theta_{23}$ is predicted to be 
somewhat on the lower side of the 3$\sigma$ allowed region. This 
can be understood by looking at the left panel plots of 
figure\ref{gsNH}, where one can see that the one-loop corrected 
results prefer lower values of $c^2_\mu$ and higher values of
$c^2_\tau$. Obviously, this gives smaller $\sin^2\theta_{23}$. On the
other hand, the tree level analysis prefers higher values of $c^2_\mu$
and both lower and higher values of $c^2_\tau$. This gives rise to large 
as well as small values of $\sin^2\theta_{23}$. 

\FIGURE{\epsfig{file=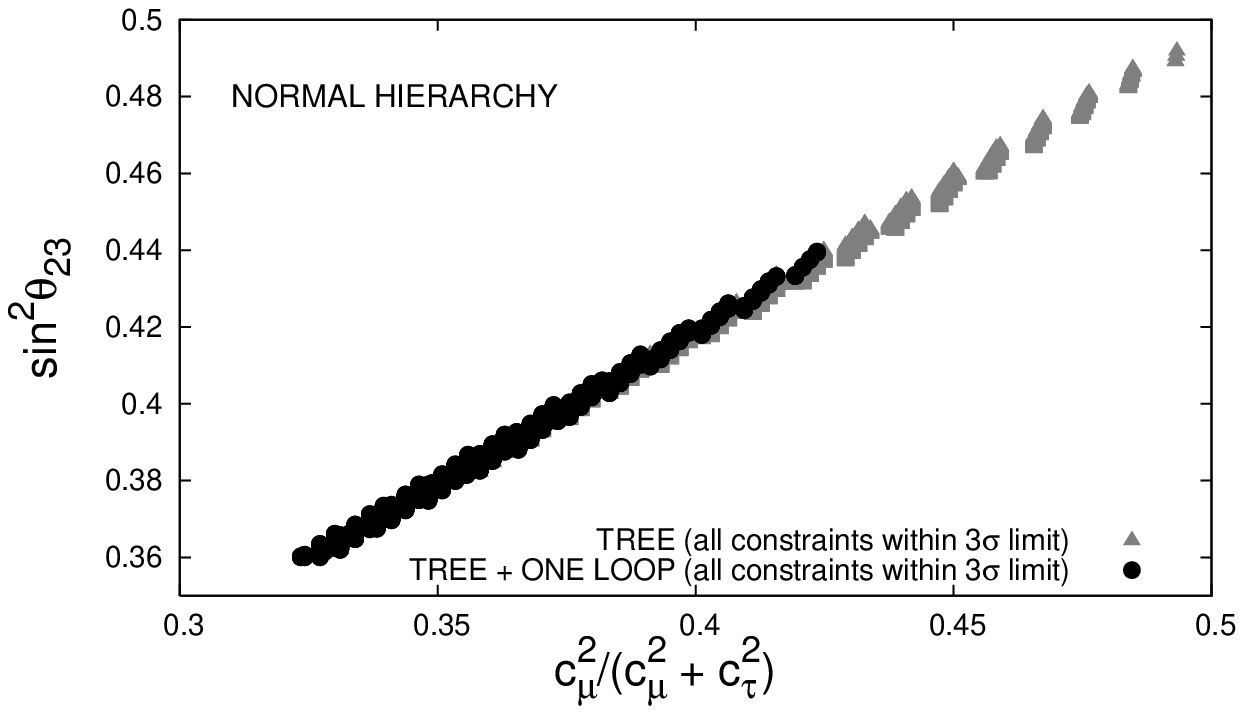,height=4.00cm} 
\vspace{0.2cm}
\epsfig{file=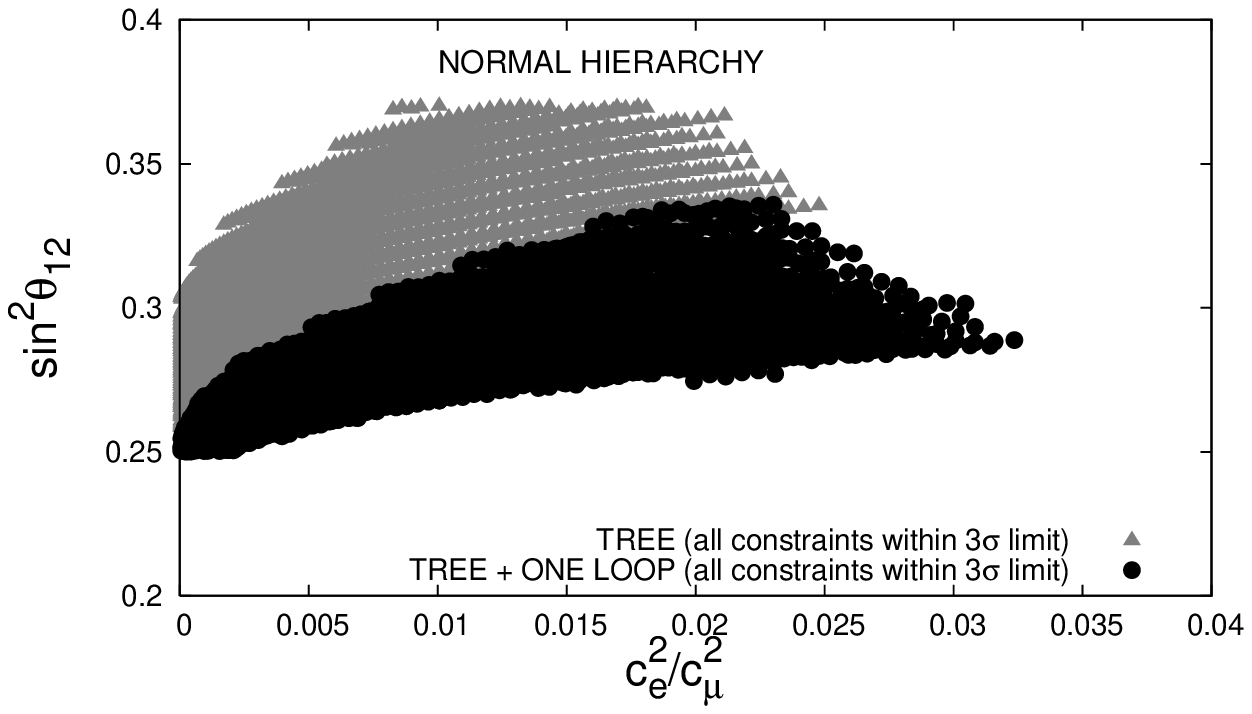,height=4.00cm} 
\vspace{0.2cm}
\epsfig{file=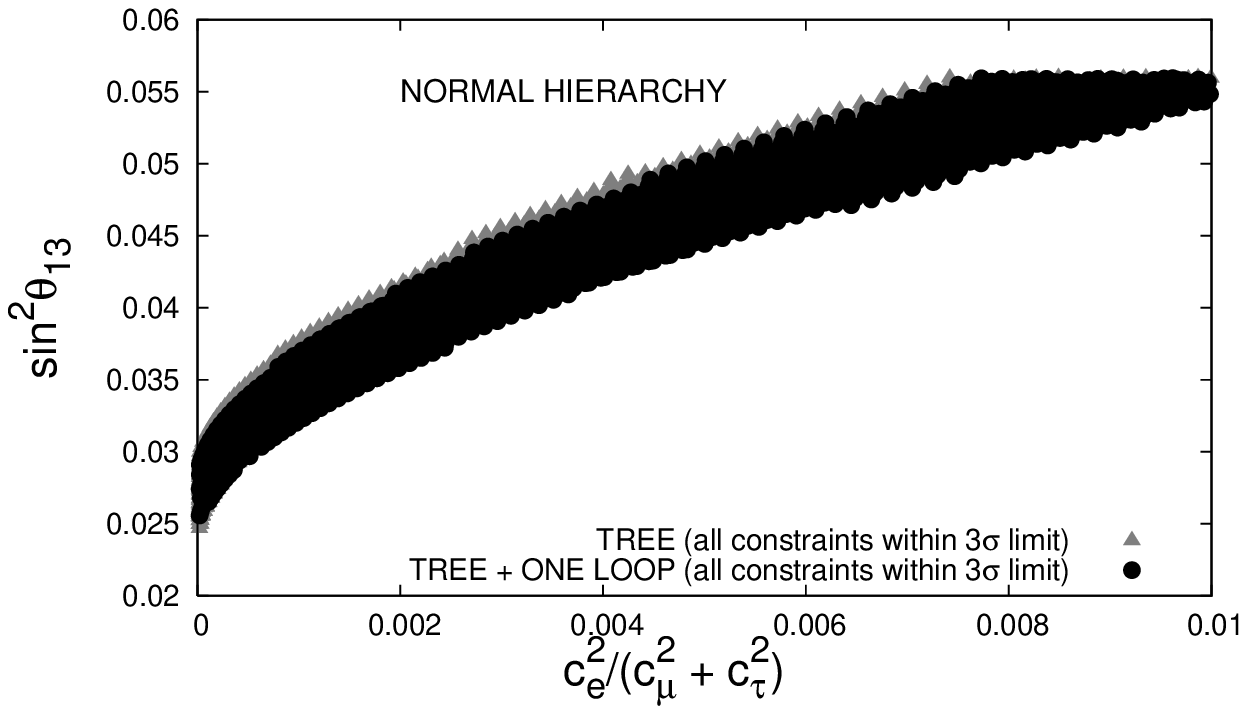,height=4.00cm} 
\vspace{0.2cm}
\caption{Variation of $\rm{sin}^2\theta_{23}$ with
$\frac{c^2_\mu}{(c^2_\mu + c^2_\tau)}$, $\rm{sin}^2\theta_{12}$ with
$\frac{c^2_e}{c^2_\mu}$, $\rm{sin}^2\theta_{13}$ with
$\frac{c^2_e}{(c^2_\mu + c^2_\tau)}$ for normal hierarchy of light
neutrino masses. Other parameter choices are discussed in the text.}
\label{mixing-GS-OS}}

If one looks at the plot of $\sin^2\theta_{13}$ in figure \ref{mixing-GS-OS}, 
then it is evident that the amount $\nu_e$ flavour in the heaviest state 
$(\nu_3)$ decreases a little bit with the inclusion of one-loop corrections 
for a fixed value of the quantity $\frac{c^2_e}{(c^2_\mu + c^2_\tau)}$. Very 
small $\sin^2\theta_{13}$ demands $c^2_e \ll c^2_\mu, ~c^2_\tau$. This feature
is also consistent with the plots in figure\ref{gsNH}. The correlation of 
$\sin^2\theta_{12}$ with the ratio $c^2_e/c^2_\mu$ is not very sharp as expected 
from the discussion given above. However, a large $\theta_{12}$ mixing angle 
requires a larger value of this ratio. The effect of one-loop correction is
more pronounced in this case and predicts a smaller value of $\sin^2\theta_{12}$ 
compared to the tree level result. There is no specific correlation of the mixing 
angles with the quantities $a^2_i$ and we do not show them here. 

\subsection{Inverted hierarchy}
\label{Inverted hierarchy}

In this subsection we perform a similar numerical analysis for the
inverted hierarchical scheme of three light neutrino masses. 
Recall that for the inverted hierarchical pattern of light neutrino 
masses, the absolute values of the mass eigenvalues are such that 
$m_2 > m_1\gg m_3$. Thus the solar and the atmospheric mass squared 
differences are defined as $\Delta m^2_{atm} = m^2_1 - m^2_3$ and 
$\Delta m^2_{solar} = m^2_2 - m^2_1$. In order to generate such a mass 
pattern, the choices of neutrino Yukawa couplings $Y_\nu^{ii}$ and the
left-handed sneutrino VEVs $v^\prime_i$ are shown in table \ref{table2}. 
However, these are just sample choices and other choices also exist 
as we will see during the course of this discussion. The choices of 
other parameters are shown in table \ref{table1}. The effect of 
one-loop corrections to the mass eigenvalues are such that the absolute 
values of masses $m_3$ and $m_1$ become smaller whereas $m_2$ grows in 
magnitude. This effect of increasing the absolute value of $m_2$ while 
decreasing that of $m_1$ makes it extremely difficult to account for the 
present 3$\sigma$ limits on $\Delta m^2_{solar}$. 

\FIGURE{\epsfig{file=Figures1/masssqesgs1sIH.eps,height=4.55cm} 
\vspace{0.2cm}
\epsfig{file=Figures1/masssqesos1sIH.eps,height=4.55cm} 
\vspace{0.2cm}
\epsfig{file=Figures1/masssqesgs2sIH.eps,height=4.55cm} 
\vspace{0.2cm}
\epsfig{file=Figures1/masssqesos2sIH.eps,height=4.55cm} 
\vspace{0.2cm}
\epsfig{file=Figures1/masssqesgs3sIH.eps,height=4.55cm} 
\vspace{0.2cm}
\epsfig{file=Figures1/masssqesos3sIH.eps,height=4.55cm}
\caption{Neutrino mass squared values ($m^2_i$) vs $\frac{c^4_i}{M^2}$ 
(left panel) and vs $\frac{a^4_i}{m^2_{\nu^c}}$ (right panel) plots for 
the {\it{inverted hierarchical}} pattern of light neutrino masses, $i = e,
 \mu, \tau$.}
\label{numsqIH}}

Typical mass spectra are shown in figure \ref{numsqIH}. Once again 
note that a particular model parameter has been varied while the others 
are fixed at values given in tables \ref{table1} and \ref{table2}. As it 
is evident from these plots, the masses $m_1$ and $m_3$ are controlled 
mainly by the parameters $a^2_i/m_\nu^c$, whereas the mass $m_2$ is 
controlled by the seesaw parameters $c^2_i/M$ though there is a small 
contribution coming from $a^2_i/m_\nu^c$ as well. 

Let us now turn our attention to the variation of $|\Delta m^2_{atm}|$ 
and $\Delta m^2_{solar}$ with $c^4_i/M^2$ and $a^4_i/{m^2_{\nu^c}}$ 
shown in figure \ref{gsIH} and figure \ref{osIH}. For our numerical analysis, 
we have set the scale of $m_3$ as  $|m_3|<0.011~\rm{eV}$. The left panel 
in figure \ref{gsIH} shows that $|\Delta m^2_{atm}|$ increases 
with $c^4_{\mu,\tau}/M^2$ and decreases with $c^4_e/M^2$. This is 
essentially the behaviour shown by $m^2_1$ with the variation of 
$c^4_i/M^2$. Similar behaviour is obtained for the one-loop corrected 
$\Delta m^2_{atm}$. The decrease in the one-loop corrected result 
compared to the tree level one is due to the splitting in $m^2_1$ 
value as shown in figure\ref{numsqIH}. 

\FIGURE{\epsfig{file=Figures1/atmsqvsgs1sIH.eps,height=4.55cm} 
\vspace{0.2cm}
\epsfig{file=Figures1/solsqvsgs1sIH.eps,height=4.55cm} 
\vspace{0.2cm}
\epsfig{file=Figures1/atmsqvsgs2sIH.eps,height=4.55cm} 
\vspace{0.2cm}
\epsfig{file=Figures1/solsqvsgs2sIH.eps,height=4.55cm} 
\vspace{0.2cm}
\epsfig{file=Figures1/atmsqvsgs3sIH.eps,height=4.55cm} 
\vspace{0.2cm}
\epsfig{file=Figures1/solsqvsgs3sIH.eps,height=4.55cm}
\caption{Atmospheric and solar mass squared differences $(|\Delta
  m^2_{atm}|,~\Delta m^2_{solar})$ vs $\frac{c_i^4}{M^2}$ plots for
  the {\it{inverted hierarchical}} pattern of light neutrino masses
  with $i = e, \mu, \tau$. Colour specification is same as described 
  in the context of figure \ref{gsNH}. Choices of other parameters are 
  discussed in the text.}
\label{gsIH}}

The variation of $\Delta m^2_{solar}$ with $c^4_i/M^2$ can be understood 
in a similar manner by looking at figure\ref{numsqIH}. As explained earlier, 
in the case of $\Delta m^2_{solar}$, the one-loop corrected result is larger 
compared to the tree level one. The range of parameters satisfying all the 
three flavour global neutrino data are shown by the fewer dark points on 
the plots. Note that the increase of $\Delta m^2_{solar}$ at the one-loop 
level is such that we do not even see any allowed range of parameters when 
looking at the variation with respect to $c^4_{e,\tau}/M^2$. Once again, the 
behaviour of $|\Delta m^2_{atm}|$ and $\Delta m^2_{solar}$ with the change 
in the parameters $a^4_i/m^2_{\nu^c}$ (shown in figure\ref{osIH}) can be 
explained by looking at the right panel plots of figure\ref{numsqIH}. 
  
\FIGURE{\epsfig{file=Figures1/atmsqvsos1sIH.eps,height=4.55cm} 
\vspace{0.2cm}
\epsfig{file=Figures1/solsqvsos1sIH.eps,height=4.55cm} 
\vspace{0.2cm}
\epsfig{file=Figures1/atmsqvsos2sIH.eps,height=4.55cm} 
\vspace{0.2cm}
\epsfig{file=Figures1/solsqvsos2sIH.eps,height=4.55cm} 
\vspace{0.2cm}
\epsfig{file=Figures1/atmsqvsos3sIH.eps,height=4.55cm} 
\vspace{0.2cm}
\epsfig{file=Figures1/solsqvsos3sIH.eps,height=4.55cm}
\caption{Atmospheric and solar mass squared differences $(|\Delta
  m^2_{atm}|,~\Delta m^2_{solar})$ vs $a^4_i/m^2_{\nu^c}$ plots for
  the {\it{inverted hierarchical}} pattern of light neutrino masses
  with $i = e, \mu, \tau$. Colour specification is same as described 
  in the context of figure \ref{gsNH}. Choices of other parameters are 
  discussed in the text.}
\label{osIH}}

We have also investigated the nature of variation of $|\Delta m^2_{atm}|$
and $\Delta m^2_{solar}$ with $\varepsilon^2_i$, the squared effective 
bilinear $R_P$-violating parameters. $|\Delta m^2_{atm}|$ was found to 
increase with $\varepsilon^2_i$ (the increase is sharper for $\varepsilon^2_1$), 
whereas $\Delta m^2_{solar}$ initially increases very sharply with $\varepsilon^2_i$ 
(particularly for $\varepsilon^2_1$ and $\varepsilon^2_2$) and then becomes 
flat. In the one-loop corrected results we do not find any range of values for
parameters where the neutrino data are satisfied. 

The variation of mass squared differences with $\lambda$ and 
$\kappa$ have also been analyzed. The variation of
$|\Delta m^2_{atm}|$ and $\Delta m^2_{solar}$ with $\lambda$ 
and $\kappa$ are found to be opposite to those of normal hierarchical
scenario. The one-loop corrected results do not show any allowed ranges 
of $\lambda$ and $\kappa$ (for the chosen values of other parameters) 
where the neutrino data can be satisfied.

The $\tan\beta$ dependence of $|\Delta m^2_{atm}|$ and $\Delta m^2_{solar}$ 
is shown in figure\ref{tanbetaIH}. One can see from these two figures that
$|\Delta m^2_{atm}|$ initially increases and then starts decreasing at a 
value of $\tan\beta$ around $10$. On the other hand, $\Delta m^2_{solar}$ 
initially decreases and then starts increasing around the same value of 
$\tan\beta$. Note that the one-loop corrected result for $|\Delta m^2_{atm}|$
is lower than the corresponding tree level result for $\tan\beta < 10$ whereas
the one-loop corrected result for $\Delta m^2_{solar}$ is lower than the 
corresponding tree level result for $\tan\beta > 10$. For the chosen values 
of other parameters we see that the one-loop corrected analysis does not 
provide any value of $\tan\beta$ where the neutrino data can be satisfied.

\FIGURE{\epsfig{file=Figures1/atmsqvstanbIH.eps,height=4.55cm} 
\vspace{0.2cm}
\epsfig{file=Figures1/solsqvstanbIH.eps,height=4.55cm} 
\caption{$|\Delta m^2_{atm}|,~\Delta m^2_{solar}$ vs $\rm{tan}\beta$
  plots for the {\it{inverted hierarchical}} pattern of light neutrino
  masses. Choices of other parameters are discussed in the text.}
\label{tanbetaIH}}

We conclude the discussion on inverted hierarchy by addressing
the dependence of neutrino mixing angles with the relevant parameters. 
In figure\ref{mixing-GS-OS-IH-B} we show the variation of the neutrino 
mixing angles with the same set of parameters as chosen for the normal
hierarchical scenario. We notice that for inverted hierarchy 
the quantity $\sin^2\theta_{23}$ decreases with increasing 
$\frac{c^2_\mu}{(c^2_\mu + c^2_\tau)}$ which is just opposite to that 
of the normal hierarchy (see, figure\ref{mixing-GS-OS}). Nevertheless, the
correlation of $\sin^2\theta_{23}$ with $\frac{c^2_\mu}{(c^2_\mu + c^2_\tau)}$
is as sharp as in the case of normal hierarchy. A similar feature is obtained
for the variation with $\frac{a^2_\mu}{(a^2_\mu + a^2_\tau)}$.

On the other hand, the correlations of $\sin^2\theta_{12}$ with 
$\frac{c^2_e}{c^2_\mu}$ and $\frac{a^2_e}{a^2_\mu}$ and the 
correlations of $\sin^2\theta_{13}$ with 
$\frac{c^2_e}{(c^2_\mu + c^2_\tau)}$ and 
$\frac{a^2_e}{(a^2_\mu + a^2_\tau)}$ are not very sharp and
we do not show them here. There are allowed values of relevant parameters
where all neutrino data can be satisfied. Remember that, for 
the plots with $c_i$s, we varied all the $c_i$s simultaneously, keeping 
the values of $a_i$s fixed at the ones determined by the parameters in 
table \ref{table2}. Similarly, for the variation of $a_i$s, the quantities 
$c_i$s were fixed. The inclusion of one-loop corrections restrict the 
allowed values of parameter points significantly compared to the tree level 
results. 

\FIGURE{\epsfig{file=Figures1/sinsq23GSbIH.eps,height=4.75cm} 
\vspace{0.2cm}
\epsfig{file=Figures1/sinsq23OSaIH.eps,height=4.75cm} 
\caption{Variation of $\sin^2\theta_{23}$ with
  $\frac{c^2_\mu}{(c^2_\mu + c^2_\tau)}$ and 
  $\frac{a^2_\mu}{(a^2_\mu + a^2_\tau)}$ 
  for inverted hierarchy of light neutrino masses. Choices of other
  parameters are discussed in the text.}
\label{mixing-GS-OS-IH-B}}

\subsection{Quasi-degenerate spectra}
\label{Quasi degenerate spectra}

The discussion on the light neutrino mass spectrum remains incomplete
without a note on the so-called ``quasi-degenerate'' scenario. A truly 
degenerate scenario of three light neutrino masses is, however, inconsistent 
with the oscillation data (see eq.(\ref{oscillation-data})). 
Hence, the quasi-degenerate scenario of light neutrino masses is defined 
in such a way that in this case all the three individual neutrino masses are 
much larger compared to the atmospheric neutrino mass scale. Mathematically,
one writes $m_1 \approx m_2 \approx m_3 \gg \sqrt{|\Delta m^2_{atm}|}$. 
Obviously, the oscillation data suggest that even in such a situation there 
must be a mild hierarchy among the degenerate neutrinos.

In this section we have shown that the huge parameter space of $\mu\nu$SSM 
always leaves us with enough room to accommodate quasi-degenerate spectrum.
For our numerical analysis, we called a set of light neutrino masses to be
quasi-degenerate if the lightest among them is greater than 0.1 eV. We choose
two sets of sample parameter points which are shown in table \ref{table3}
(values of other parameters are same as in table \ref{table1}). For these 
two sets of neutrino Yukawa couplings ($Y_\nu^{ii}$) and the left-handed 
sneutrino VEVs ($v^\prime_i$) we observe the following patterns of light
neutrino masses at the tree level   

\noindent (i) Quasi-degenerate-I: $m_3 \gsim m_2 \gsim m_1 \gg 
\sqrt{|\Delta m^2_{atm}|},$
\\
(ii) Quasi-degenerate-II: $m_2 \gsim m_1 \gsim m_3 \gg \sqrt{|\Delta m^2_{atm}|}.$ 

\noindent For case (i), we have varied the parameters around the values in 
table \ref{table3} and identified a few extremely fine-tuned points in the 
parameter space where either the tree level or the one-loop corrected result 
is consistent with the three flavour global neutrino data. Two representative
spectrum as function of $\frac{c^4_e}{M^2}$ and $\frac{a^4_e}{m^2_{\nu^c}}$ 
are shown in figure \ref{numsqQD}.

\FIGURE{\epsfig{file=Figures1/masssqesgs1sQD.eps,height=5.00cm} 
\vspace{0.2cm}
\epsfig{file=Figures1/masssqesos1sQD.eps,height=5.00cm} 
\caption{Neutrino mass squared values ($m^2_i$) vs $\frac{c^4_e}{M^2}$ 
(left panel) and vs $\frac{a^4_e}{m^2_{\nu^c}}$ (right panel) plots for 
the {\it{quasi-degenerate}} pattern of light neutrino masses. 
}
\label{numsqQD}}

\TABLE[t]
{\begin{tabular}{|c|c|c|c|c|c|c|}
\hline
 & \multicolumn{3}{c|}{$Y^{ii}_\nu \times 10^7$} 
& \multicolumn{3}{c|}{$v'_i \times 10^5 (\rm{GeV})$ } \\
\cline{2-7}
 & $Y^{11}_\nu$ & $Y^{22}_\nu$ & $Y^{33}_\nu$ & $v'_1$ 
& $v'_2$ & $v'_3$ \\
\hline
\text{Quasi-degenerate-I} & 19.60 & 19.94 & 19.99 & 9.75 & 10.60 & 11.83\\
\hline
\text{Quasi-degenerate-II} & 18.50  & 18.00 & 18.00 & 9.85  & 10.50 & 10.10\\
\hline
\end{tabular}
\caption{Values of neutrino Yukawa couplings and left handed sneutrino VEVs,
  used as sample parameter points to obtain quasi-degenerate light neutrino 
  spectrum. Around these values, the corresponding parameters were
  varied for the plots shown in figure \ref{numsqQD}. \label{table3}}}

As mentioned earlier, one can play with the model parameters and 
obtain a spectrum with a different ordering of masses 
termed as ``Quasi-degenerate-II" in table \ref{table3}. However, for
such an ordering of masses, we found that it was rather impossible
to find any region of parameter space where the one-loop corrected result
satisfies all the constraints on neutrino masses and mixing. 
Nevertheless, we must emphasize here that it is not a 
completely generic conclusion and for other choices of soft SUSY breaking 
and other parameters it could be possible to have a spectrum like that 
shown in ``Quasi degenerate II" with neutrino constraints satisfied 
even at the one-loop level. On the other hand, there exists regions where 
neutrino data are satisfied at the tree level with this ordering of masses.


\section{Summary and Conclusion}
\label{Summary and conclusion}
In this paper, we have performed a systematic study of neutrino masses
and mixing in a non-minimal extension of the MSSM, known as
$\mu\nu$SSM, with the complete set of one-loop radiative
corrections, over and above the effects from the seesaw mechanism 
(of Types I and III). A set of right chiral neutrino superfields, 
introduced in this model provide a solution
to the $\mu$-problem. Lepton number is broken by one $\Delta L$ = 1
term comprising the right-chiral neutrinos and the Higgs superfields
and another $\Delta L$ = 3 term involving only the right chiral neutrinos
in the superpotential. The corresponding soft SUSY breaking terms in the 
scalar potential together with the F-term contribution from the 
superpotential, induce VEVs of sneutrinos of both right and left chirality.
The right chiral neutrinos, together with neutralinos, are instrumental 
in the generation of light neutrino masses at the tree level.

In \cite{Ghosh-Roy}, where a tree-level analysis of neutrino masses
and mixing was carried out, it was shown that tree level masses could be
generated in this fashion for all three light neutrinos, even with a flavour 
diagonal structure of neutrino Yukawa couplings ($Y_\nu^{ij}$). In this work 
we have improved the analysis with the inclusion of one-loop radiative 
corrections, for various patterns of light neutrino masses, namely, 
the normal, inverted and quasi-degenerate spectra, 
Attempts have been made to identify regions in the SUSY parameter space, which
can accommodate the three patterns in turn. Our analysis clearly shows that 
the multi-dimensional parameter space of $\mu\nu$SSM leaves enough room 
to accommodate all the diverse mass hierarchies of the three active light neutrinos.

As a prerequisite of the loop calculation, we have derived the entire set of 
Feynman rules necessary for our analytical and numerical studies. The set of 
rules will be of immense importance, should one want to look for signatures of 
this model at the Large Hadron Collider (LHC), specifically through decays of the
new particles of this model into SM particles. Approximate analytical
forms of the entries of the expansion matrix `$\xi$' have also been
provided including all three generations of right handed neutrinos. A
handful of relations within the four-component weak and mass eigenbasis 
for scalar and fermion fields of $\mu\nu$SSM have also been worked out.

The correlation of neutrino mass squared differences and mixing angles 
with relevant model parameters, consistent with the EWSB conditions, 
have been studied in detail for different mass hierarchies. 
The allowed regions of the parameter space are found to be rather seriously 
affected by the one-loop contributions. We have also observed that the 
globally fitted neutrino data can be accommodated into the $\mu\nu$SSM for 
various compositions of the Lightest Supersymmetric Particle (LSP), once one 
starts playing with the parameters of the model. Although for the present study, we
adhered to a Bino dominated LSP scheme, we numerically verified
the possibility of having a light right-chiral neutrino (i.e. 
{\it{singlino}}) like LSP, compatible with neutrino data. This, by
itself, is an interesting scenario in the sense that the singlino LSP
offers a scope to probe the right-handed neutrino mass-scale at the
LHC. It would also be interesting to perform an explicit radiative
correction to the heavy neutralinos. A dedicated analysis for this is
beyond the scope of this paper, and a future publication with
comprehensive discussion of all these issues may be well
anticipated. In conclusion, one-loop radiative corrections to the
neutrino masses and mixing angles for $\mu\nu$SSM are capable of
substantially altering the tree level analysis.

\vspace{1cm}

\noindent {\bf Acknowledgments} 

\noindent We thank Utpal Chattopadhyay, Debajyoti Choudhury, Debottam Das,
Anindya Datta, Aseshkrishna Datta, Dilip Kumar Ghosh, Palash Baran Pal,
Subhendu Rakshit and Sreerup Raychaudhuri for helpful discussions. PG
would like to thank the Council of Scientific and
Industrial Research, Government of India, for a Senior Research Fellowship. 
PD is supported through the Gottfried Wilhelm Leibniz Program by the Deutsche
Forschungsgemeinschaft (DFG). The work of PD and BM was partially supported
by funding available from the Department of Atomic Energy, Government of India,
for the Regional Centre for Accelerator-based Particle Physics (RECAPP), 
Harish-Chandra Research Institute. These two authors also acknowledge
the hospitality of the Department of Theoretical Physics of the Indian Association
for the Cultivation of Science (IACS), while this work was in progress. PG and SR 
wish to thank RECAPP for hospitality during a part of the investigation. 
Computational work for this study was carried out using the cluster computing 
facility at the Department of Theoretical Physics, IACS.

\appendix


\section{Minimization equations}\label{Minimization equations}
The minimization equations with respect to the VEVs
$v^c,~v^{\prime}_i,~v_2,~v_1$ are given below.

\beq
2{\sum_{j}} {{u^{ij}_c}} {\zeta^{j}} + \sum_{k} Y^{ki}_{\nu} {r^{k}_c} {v_2^2}
+\sum_{j} (m^2_{\widetilde{\nu}^c})^{ji} 
{v^c_j}+ {\rho^i \eta} + {\mu} {\lambda^i v_2^2}+(A_x x)^{i} =0,
\label{Minim1}
\eeq
\beq
{\sum_{j}} {Y_{\nu}}^{ij} {v_2} {\zeta^{j}}  +\sum_{j} (m^2_{\widetilde{L}})^{ji} {v'_j}+\sum_{j} (A_{\nu} 
Y_{\nu})^{ij} {v^c_j} v_2 +{\gamma_g}{\xi_{\upsilon}}{v'_i}
+{r^i_c} {\eta} =0,
\label{Minim2}
\eeq
\beq
{\sum_{j}}{\rho^{j}}{\zeta^{j}} + {{\sum_{i}} {r^i_c}^2 v_2}+ {\sum_{i}}({A_{\nu}Y_{\nu}})^{ij} 
{v'_i} {v^c_j}-\sum_{i}({A_{\lambda} {\lambda}})^i {v^c_i} v_1 + {m^2_{H_2}}v_2 +{\mu^2} v_2
-{\gamma_g}{\xi_{\upsilon}} {v_2} 
= 0,
\label{Minim3}
\eeq
\beq
-{\sum_{j}}{\lambda^j}v_2 {\zeta^{j}}- {\mu} {\sum_{j}} {r^j_c} {v'_j} -\sum_{i}({A_{\lambda} 
{\lambda}})^i {v^c_i} v_2 + {m^2_{H_1}}v_1 +{\gamma_g}{\xi_{\upsilon}} {v_1} + 
{\mu^2} v_1 = 0,
\label{Minim4}
\eeq
where
\bea
(A_x x)^{i} &=& \sum_{j} (A_{\nu} Y_{\nu})^{ji} {v'_j} v_2 + \sum_{j,k} 
({A_\kappa} {\kappa})^{ijk} {v^c_j} {v^c_k} - (A_{\lambda} {\lambda})^i v_1 v_2,  \nonumber \\
{r^{i}_c} &=& \varepsilon^i = {\sum_{j}Y^{ij}_{\nu} {v^c_{j}}}, ~~{r^{i}} = {\sum_{j}Y^{ij}_{\nu} {v'_{j}}}, ~~{u^{ij}_c} = {\sum_{k}}\kappa^{ijk}{v^c_k}, \nonumber \\
\zeta^{j} &=& \sum_{i} 
{u^{ij}_c} {v^c_i} + {r^j} v_2 - 
{\lambda}^j v_1 v_2 ,  ~~\mu = \sum_{i}\lambda^{i} {v^c_i},\nonumber \\ 
\eta &=&\sum_{i} {r^i_c}{v'_i}  - {\mu} v_1 , ~~\rho^{i} = {r^i} - \lambda^i v_1,\nonumber \\
\gamma_{g} &=& \frac{1}{4}({g_1^2 + g_2^2}), ~~\xi_{\upsilon} ={\sum_{i} {v'^2_i} + v_1^2 -v_2^2}.  \nonumber \\
\label{Abbrevations}
\eea
In deriving the above equations, it has been assumed that 
${\kappa}^{ijk}$, $({{A_\kappa} {\kappa}})^{ijk}$,  $Y^{ij}_{\nu}$,
$(A_{\nu} Y_{\nu})^{ij}$, $(m^2_{\widetilde{\nu}^c})^{ij}$, 
$(m^2_{\widetilde{L}})^{ij}$ are all symmetric in $i,j,k$. 


\section{Details of expansion matrix $\xi$}\label{Details of expansion matrix}
In this appendix the entries of the expansion matrix $\xi$ are given in details

\bea\label{expansion-parameter-terms}
& &\xi_{i1} \approx \frac{\sqrt{2} g_1 \mu m^2_{\nu^c} M_2 A}{12 D}b_i, \nonumber \\
& &\xi_{i2} \approx -\frac{\sqrt{2} g_2 \mu m^2_{\nu^c} M_1 A}{12 D}b_i, \nonumber \\
& &\xi_{i3} \approx -\frac{m^2_{\nu^c} M^{\prime}}{2 D} \left\{ {\left(\lambda v_2 v^2 - 4 \mu A {\frac{M}{v_2}}\right)}a_i + {m_{\nu^c} v_2 v^c}b_i-{3\lm \left(\lm v_1 v^2-2 m_{\nu^c} v^c v_2\right) }c_i\right\}, \nonumber \\
& &\xi_{i4} \approx -\frac{m^2_{\nu^c} M^{\prime}}{2 D} \left\{ {\lm v_1 v^2}a_i + {m_{\nu^c} v_1 v^c}b_i+{3\lm^2 v_2 v^2}c_i\right\}, \nonumber \\
& &\xi_{i,4+i} \approx \frac{m_{\nu^c} M^{\prime}}{2 D}\left\{ {2 \lm \left( \lm v^4 (1-\frac{1}{2}{\rm{sin}^2{2\beta}}) + \frac{m_{\nu^c}}{2} v^c v^2 {\rm{sin}{2\beta}}  + A v^2 {\rm{sin}{2\beta}} - 4 \mu M A\right)}a_i - {\mu m_{\nu^c} v^2 {\rm{cos}{2\beta}}}b_i\right\}, \nonumber \\
& &\xi_{16} \approx \xi_{17} \approx -\frac{m_{\nu^c}  M^{\prime}}{2 D}\left\{ {\lm \left( \lm {v^4} - 4 \mu M A\right)}a_1 + \frac{\mu m_{\nu^c} v^2}{3}b_1 - {2 \lm \mu m_{\nu^c} v^2_2}c_1\right\}, \nonumber \\
& &\xi_{25} \approx \xi_{27} \approx -\frac{m_{\nu^c}  M^{\prime}}{2 D}\left\{ {\lm \left( \lm {v^4} - 4 \mu M A\right)}a_2 + \frac{\mu m_{\nu^c} v^2}{3}b_2 - {2 \lm \mu m_{\nu^c} v^2_2}c_2\right\}, \nonumber \\
& &\xi_{35} \approx \xi_{36} \approx -\frac{m_{\nu^c}  M^{\prime}}{2 D}\left\{ {\lm \left( \lm {v^4} - 4 \mu M A\right)}a_3 + \frac{\mu m_{\nu^c} v^2}{3}b_3 - {2 \lm \mu m_{\nu^c} v^2}c_3\right\}, 
\eea
where
\bea
& &a_i = Y_{\nu}^{ii} v_2, ~b_i = (Y_{\nu}^{ii} v_1 + 3 {\lambda} {v'_i}),~c_i = {v'_i}, \nonumber \\ 
& &m_{\nu^c} = 2 \kp v^c, ~\mu = 3 \lambda v^c,~A = ({\kappa}{v^c}^2 + {\lambda} v_1 v_2),\nonumber \\
& &v_2 = v {\rm{sin}{\beta}},~v_1 = v \rm{cos}{\beta}, ~D = Det\left[M_{7\times7}\right],\nonumber \\
& &\frac{1}{M} = \frac{g^2_1}{M_1} +\frac{g^2_2}{M_2}, ~M^{\prime} = \frac{M_1 M_2}{M}, 
\label{specifications-2}
\eea
with ${i} = {e,\mu,\tau} ~\equiv{1,2,3}$.

 \section{Scalar mass squared matrices}\label{Scalar mass squared matrices}
The superpotential of the $\mu \nu$SSM violates $R_P$ through lepton($L$) number violation. 
This allows the Higgses (having zero lepton number) to mix with the 
sleptons (having non-zero lepton number). Hence, the neutral (both $CP$-odd and $CP$-even) 
and charged Higgs mass squared matrices are enlarged to $8\times8$, considering 
all three slepton generation. The independent entries of the $CP$-odd, $CP$-even and the 
charged scalar mass squared matrices were derived using eqs.(\ref{Minim3}), (\ref{Minim4}), 
and eq.(\ref{Abbrevations}). Details of each of these matrix elements were given in 
ref. \cite{Ghosh-Roy}, hence we do not repeat them here. However, we give the expressions 
for scalar quark (squark) mass squared matrices.

In this appendix we present the relevant details of various scalar mass squared 
matrices required for the Feynman rules. The scalar sector of this model have also 
been addressed in ref.\cite{munoz-lopez-2,Ghosh-Roy,Porod-Bartl,munoz-lopez-3}. 


\subsection{CP-odd neutral mass squared matrix}
\label{CP-odd neutral mass squared matrix}

In the weak interaction basis $\Phi^T_{P}=(H^0_{1{\mathcal I}},H^0_{2{\mathcal I}}, {\widetilde \nu}^c_{n{\mathcal I}},{\widetilde \nu}_{n{\mathcal I}})$, the pseudoscalar mass term in the Lagrangian is of the form
\beq\label{pseudoscalar_Lagrangian}
{\mathcal{L}_{pseudoscalar}^{mass}} = {\Phi_{P}^T} {M}^2_{P} {\Phi_{P}},
\eeq
where ${M}^2_{P}$ is an $8\times8$ symmetric matrix. The mass eigenvectors are defined as
\bea
P^0_\alpha = R^{P^0}_{\alpha \beta} \Phi_{P_\beta},
\label{pseudoscalar-mass-basis}
\eea
with the diagonal mass matrix
\bea
(\mathcal{M}^{diag}_{P})^2_{\alpha \beta} = R^{P^0}_{\alpha \gamma} {M}^2_{P_{\gamma \delta}} R^{P^0}_{\beta \delta}.
\label{pseudoscalar-diagonal-mass-matrix}
\eea


\subsection{CP-even neutral mass squared matrix}
\label{CP-even neutral mass squared matrix}
In the flavour basis or weak interaction basis $\Phi^T_{S}=({H^0_{1{\mathcal R}}},{H^0_{2{\mathcal R}}},
{{\widetilde{\nu}}^c_{n{\mathcal R}}},{{\widetilde{\nu}_{n{\mathcal R}}}})$, the scalar mass term in the Lagrangian is of the form

\beq\label{scalar_Lagrangian}
{\mathcal{L}_{scalar}^{mass}} = {\Phi_{S}^T} {M}^2_{S} {\Phi_{S}},
\eeq
where ${M}^2_{S}$ is an $8\times8$ symmetric matrix. The mass eigenvectors are 
\beq
S^0_\alpha = R^{S^0}_{\alpha \beta} \Phi_{S_\beta},
\label{scalar-mass-basis}
\eeq
with the diagonal mass matrix
\beq
(\mathcal{M}^{diag}_{S})^2_{\alpha \beta} = R^{S^0}_{\alpha \gamma} {M}^2_{S_{\gamma \delta}} R^{S^0}_{\beta \delta}.
\label{scalar-diagonal-mass-matrix}
\eeq


\subsection{Charged scalar mass squared matrix}
\label{Charged scalar mass squared matrix}

In the weak basis $\Phi^{+^T}_{C}=({H^+_{1}}, {H^+_{2}}, 
{{\widetilde{e}}^+_{Rn}} ,{{\widetilde{e}}^+_{Ln}}),$ basis the charged scalar mass term in the Lagrangian is of the form
\beq\label{charged-scalar_Lagrangian}
{\mathcal{L}_{charged~scalar}^{mass}} = {\Phi_{C}^{-^T}} {M}^2_{C^{\pm}} 
{\Phi_{C}^{+}},
\eeq
where ${M}^2_{C^{\pm}}$ is an $8\times8$ symmetric matrix. The mass eigenvectors are 
\bea
S^{\pm}_\alpha = R^{S^{\pm}}_{\alpha \beta} \Phi^{\pm}_{C_\beta},
\label{charged-scalar-mass-basis}
\eea
with the diagonal mass matrix
\bea
(\mathcal{M}^{diag}_{C^{\pm}})^2_{\alpha \beta} = R^{S^{\pm}}_{\alpha \gamma} {M}^2_{C^{\pm}_{\gamma \delta}} R^{S^{\pm}}_{\beta \delta}.
\label{charged-scalar-diagonal-mass-matrix}
\eea

One of the eight eigenvalues of the CP-odd scalar and the charged scalar mass squared matrix is zero and corresponds to the neutral and the charged Goldstone boson, respectively. We reiterate that explicit expressions for the independent entries of 
${M}^2_{P},~{M}^2_{S} ~{\rm{and}}~{M}^2_{C^{\pm}}$ are given in ref. \cite{Ghosh-Roy}.


\subsection{Scalar quark mass squared matrix}
\label{Scalar quark mass squared matrix}

In the weak basis,
$\widetilde{u'}_i=(\widetilde{u}_{L_i},\widetilde{u}^*_{R_i})$ and 
$\widetilde{d'}_i=(\widetilde{d}_{L_i},\widetilde{d}^*_{R_i} )$, we get

\beq
\mathcal{L}^{mass}_{squark}=
     \frac{1}{2}\widetilde{u'_i}^\dag M_{\widetilde{u_{ij}}}^2\ \widetilde{u'_j}
     +\frac{1}{2}\widetilde{d'_i}^{\dag} M_{\widetilde{d_{ij}}}^2\ \widetilde{d'_j}\ ,
     \label{squark-mass-lagrangian}
\eeq
where $\widetilde{q}=(\widetilde{u'},\widetilde{d'})$. Explicitly for up and down type squarks $(\widetilde{u},\widetilde{d})$, using eq.(\ref{Abbrevations}) the entries are
\bea
(M^2_{\widetilde{u}})^{L_iL_j}&=&
    (m^2_{\widetilde{Q}})^{ij} + \frac{1}{6}(\frac{3g^2_2}{2} 
    - \frac{g_1^2}{2}){\xi_{\upsilon}}{\delta^{ij}}+ 
    \sum_n {Y_u^{in} Y_u^{jn} v_2^2}\ ,  \nonumber\\ 
(M^2_{\widetilde{u}})^{R_iR_j}&=&
    (m^2_{\widetilde{u}^c})^{ij}+ \frac{g^2_1}{3}{\xi_{\upsilon}}{\delta^{ij}}
    + \sum_n {Y_u^{ni} Y_u^{nj}v_2^2}\ ,  \nonumber\\
(M^2_{\widetilde{u}})^{L_iR_j}&=& 
    (A_u Y_u)^{ij} v_2 -Y_u^{ij} v_1 \mu  
    + Y_u^{ij} \sum_l{r^l_c v'_l}\ ,\nonumber\\
(M^2_{\widetilde{u}})^{R_iL_j}&=&
    (M^2_{\widetilde{u}})^{L_jR_i}\ ,
\label{entries-of-usquark-mass-matrix}
\eea 
and
\bea
(M^2_{\widetilde{d}})^{L_iL_j}&=&
    (m^2_{\widetilde{Q}})^{ij} - \frac{1}{6}(\frac{3g^2_2}{2} 
    + \frac{g_1^2}{2}){\xi_{\upsilon}}{\delta^{ij}}+ 
    \sum_n {Y_d^{in} Y_d^{jn} v_1^2}\ ,  \nonumber\\ 
(M^2_{\widetilde{d}})^{R_iR_j}&=&
    (m^2_{\widetilde{d}^c})^{ij} - \frac{g^2_1}{6}{\xi_{\upsilon}}{\delta^{ij}}
    + \sum_n {Y_d^{ni} Y_d^{nj}v_1^2}\ ,  \nonumber\\
(M^2_{\widetilde{d}})^{L_iR_j}&=& 
    (A_d Y_d)^{ij} v_1 -Y_d^{ij} v_2 \mu \ ,\nonumber\\
(M^2_{\widetilde{d}})^{R_iL_j}&=&
    (M^2_{\widetilde{d}})^{L_jR_i}\ .
\label{entries-of-dsquark-mass-matrix}
\eea 

For the mass eigenstate $\widetilde{\mathbf{q}}_i$ we have
\beq
\widetilde{\mathbf{q}}_i = R^{\widetilde{q}}_{ij} \widetilde{q}_j\ , 
\label{squark-mass-basis}
\eeq
with the diagonal mass matrix
\beq
 (\mathcal{M}^{\text{diag}}_{\widetilde{q}})^2_{ij} 
   = R^{\widetilde{q}}_{il}  M^2_{\widetilde{q}_{lk}} R^{\widetilde{q}}_{jk}\ .
\label{squark-diagonal-mass-matrix}
\eeq


\section{Charged fermion mass matrix}\label{Charged fermion mass matrix}


\subsection{Chargino mass matrix}
\label{Chargino mass matrix}

The lepton number $(L)$ violation allows mixing between the MSSM charginos with the charged leptons and 
thus the chargino mass matrices enhances to $5\times5$. The mixing between MSSM charginos and charged 
leptons are governed by the left handed sneutrino VEVs($v^{\prime}_i$) and neutrino Yukawa couplings. 
Both of which has to be small in order to satisfy the global neutrino data, thus this $R_P$ violating 
mixing are very small. The chargino mass matrix for the $\mu \nu$SSM have been addressed in 
ref.\cite{munoz-lopez-2,Ghosh-Roy,Porod-Bartl}.
In the weak interaction basis defined by
\bea
{\Psi^{+T}} = (-i \widetilde {\lambda}^{+}_{2}, \widetilde{H}_2^{+}, e_{R}^{+}, 
\mu_{R}^{+}, \tau_{R}^{+}), \nonumber \\
{\Psi^{-T}} = (-i \widetilde {\lambda}_{2}^{-}, \widetilde{H}_1^{-}, e_{L}^{-}, 
\mu_{L}^{-}, \tau_{L}^{-}), 
\label{chargino_basis}
\eea
the charged fermion mass term in the Lagrangian is of the form
\beq\label{chargino_mass_Lagrangian}
{\mathcal{L}_{charged}^{mass}} = -\frac{1}{2} 
\left(\begin{array}{cc}
\Psi^{+^T} & \Psi^{-^T}
\end{array}\right)
\left(\begin{array}{cc}
0_{5\times5} & m_{5\times5}^T \\ \\
m_{5\times5} & 0_{5\times5}
\end{array}\right)
\left(\begin{array}{c}
\Psi^+ \\ \\
\Psi^-
\end{array}\right).
\eeq
Here for simplicity we assume diagonal form of the charged Yukawa couplings.
The matrix $m_{5\times5}$ is given by (using eq.(\ref{Abbrevations}))
\beq\label{chargino_mass_matrix}
m_{5\times5} =
\left(\begin{array}{ccccc}
M_2 & {g_2}{v_2} & 0 & 0 & 0 \\ \\
{g_2}{v_1} & {\mu} & -{Y_{e}^{ee}}{v'_e} & -{Y_{e}^{{\mu}{\mu}}}{v'_{\mu}} &
-{Y_{e}^{{\tau}{\tau}}}{v'_{\tau}} \\ \\
{g_2}{v'_e} & -{r^e_c} & {Y_{e}^{ee}}{v_1} & 0 & 0 \\ \\
{g_2}{v'_{\mu}} & -{r^{\mu}_c} & 0 & {Y_{e}^{{\mu}{\mu}}}{v_1} & 0 \\ \\
{g_2}{v'_{\tau}} & -{r^{\tau}_c} & 0 & 0 & {Y_{e}^{{\tau}{\tau}}}{v_1}
\end{array}\right).
\eeq
The charged fermion masses are obtained by applying a bi-unitary 
transformation such that 
\beq\label{chargino_mass_eigenstate_matrixform}
U^* m_{5\times5} V^{-1} = \mathcal{M}^{\pm}_D,
\eeq
where $U^*$ and $V$ are two unitary matrices and $\mathcal{M}^{\pm}_D$ is the
diagonal matrix with non-negative entries corresponding to the
physical fermion masses. The two-component mass eigenstates are defined 
by
\bea\label{chargino_mass_eigenstate}
& &\chi^+_i= V_{ij} \Psi^+_j, \nonumber \\
& &\chi^-_i= U_{ij} \Psi^-_j, \quad i,j=1,...,5.
\eea
Nevertheless, we notice that the 13, 14, and 15 elements of the chargino mass 
matrix (eq. (\ref{chargino_mass_matrix})) are vanishing and given the orders of 
magnitude of various parameters, we also see that the values of the other 
off-diagonal entries (except for 12 and 21 elements) are very small. This 
indicates that the physical charged lepton eigenstates will have a very small 
admixture of charged higgsino and charged gaugino states. So we can very well 
assume (also verified numerically) that this mixing has very little effect on 
the mass eigenstates of the charged leptons. Thus, while writing down the 
neutrino mixing matrix, it will be justified to assume that one is  working in 
the basis where the charged lepton mass matrix is already in the diagonal form. 

\subsection{Quark mass matrix}
\label{Quark mass matrix}

The mixing matrices for up and down quarks are $3\times3$ and they are diagonalized using bi-unitary 
transformation. Entries of up and down quark mass matrices $m^u_{3\times3}~\rm{and}~m^d_{3\times3}$ are given below
\bea
(m^u_{3\times3})_{ij} &=& Y_u^{ij} v_2, \nonumber \\
(m^d_{3\times3})_{ij} &=& Y_d^{ij} v_1. 
\label{quark-mass-matrix}
\eea

The quark mass matrices are diagonalized as follows
\bea\label{quark_mass_eigenstate_matrixform}
{R^u_L}^* m^u_{3\times3} {R^u_R}^{-1} &=& \mathcal{M}^{diag}_U, \nonumber \\
{R^d_L}^* m^d_{3\times3} {R^d_R}^{-1} &=& \mathcal{M}^{diag}_D.
\eea


 \section{Feynman rules}\label{Feynman rules}
In this appendix we will study the relevant Feynman rules required for the calculations of the 
one-loop contributions to the neutralino masses. Some of the Feynman rules for this model have 
been derived in ref.\cite{Ghosh-Roy}. Feynman rules for MSSM are given in 
ref.\cite{Haber-Kane,Rosiek-loop} and in ref.\cite{Gunion-Haber-1,Franke-Fraas} for MSSM 
with singlet superfields. Feynman rules for $R_P$-violating MSSM were studied in 
ref.\cite{Hirsch-loop-FD}. The required Feynman rules are (using relations of 
appendix \ref{Some Useful Relations}, shown later) of the 
form {\it{neutralino-fermion-scalar/gauge boson}} and they are listed below.


\subsection*{Neutralino-neutralino-neutral scalar}

The Lagrangian using four component spinor notation can be written as
\beq
\mathcal{L}^{nnh}= - \frac{\widetilde g}{\sqrt{2}} \ovl{{\widetilde \chi}^0_i} (O^{nnh}_{Lijk} P_L + O^{nnh}_{Rijk} P_R) {{\widetilde \chi}^0_j} S^0_k,
\label{Neutralino-neutralino-neutral-scalar}
\eeq

where
\bea
{\widetilde g} O^{nnh}_{Lijk} = & & \eta_{j} \frac{1}{2} \left[\bRs_{k1} \left( \frac{g_2}{\sqrt{2}} {\bN}^*_{i2} \bN^*_{j3} - \frac{g_1}{\sqrt{2}} \bN^*_{i1} \bN^*_{j3} - \lm^m \bN^*_{i4} \bN^*_{j,m+4}\right)
\right. \nonumber \\
& &+ \bRs_{k2} \left( -\frac{g_2}{\sqrt{2}} {\bN}^*_{i2} \bN^*_{j4} + \frac{g_1}{\sqrt{2}} \bN^*_{i1} \bN^*_{j4} - \lm^m \bN^*_{i3} \bN^*_{j,m+4} + Y^{mn}_{\nu} \bN^*_{i,n+4} \bN^*_{j,m+7}\right)\nonumber \\
& &+ \bRs_{k,m+2} \left( Y^{mn}_{\nu} \bN^*_{i4} \bN^*_{j,n+7} - \lm^m \bN^*_{i3} \bN^*_{j4} + \kp^{mnp} \bN^*_{i,n+4} \bN^*_{j,p+4}\right)\nonumber \\
& &+ \left.
\bRs_{k,m+5} \left( \frac{g_2}{\sqrt{2}} {\bN}^*_{i2} \bN^*_{j,m+7} - \frac{g_1}{\sqrt{2}} \bN^*_{i1} \bN^*_{j,m+7} + Y^{mn}_{\nu} \bN^*_{i4} \bN^*_{j,n+4}\right) + (i\leftrightarrow j)\right],\nonumber\\
\label{nnh-L-couplings}
\eea
and
\beq
O^{nnh}_{Rijk} =(O^{nnh}_{Lijk})^*.
\label{nnh-R-couplings}
\eeq

\subsection*{Neutralino-neutralino-neutral pseudoscalar}

The Lagrangian using four component spinor notation can be written as
\beq
\mathcal{L}^{nna}= - i\frac{\widetilde g}{\sqrt{2}} \ovl{{\widetilde \chi}^0_i} (O^{nna}_{Lijk} P_L + O^{nna}_{Rijk} P_R) {{\widetilde \chi}^0_j} P^0_k,
\label{Neutralino-neutralino-neutral-pseudoscalar}
\eeq

where
\bea
{\widetilde g} O^{nna}_{Lijk} = & &\eta_{j} \frac{1}{2} \left[\bRp_{k1} \left( -\frac{g_2}{\sqrt{2}} {\bN}^*_{i2} \bN^*_{j3} + \frac{g_1}{\sqrt{2}} \bN^*_{i1} \bN^*_{j3} - \lm^m \bN^*_{i4} \bN^*_{j,m+4}\right)
\right. \nonumber \\
& &+ \bRp_{k2} \left( \frac{g_2}{\sqrt{2}} {\bN}^*_{i2} \bN^*_{j4} - \frac{g_1}{\sqrt{2}} \bN^*_{i1} \bN^*_{j4} - \lm^m \bN^*_{i3} \bN^*_{j,m+4} + Y^{mn}_{\nu} \bN^*_{i,n+4} \bN^*_{j,m+7}\right)\nonumber \\
& &+ \bRp_{k,m+2} \left( Y^{mn}_{\nu} \bN^*_{i4} \bN^*_{j,n+7} - \lm^m \bN^*_{i3} \bN^*_{j4} + \kp^{mnp} \bN^*_{i,n+4} \bN^*_{j,p+4}\right)\nonumber \\
& &+ \left.
\bRp_{k,m+5} \left( -\frac{g_2}{\sqrt{2}} {\bN}^*_{i2} \bN^*_{j,m+7} + \frac{g_1}{\sqrt{2}} \bN^*_{i1} \bN^*_{j,m+7} + Y^{mn}_{\nu} \bN^*_{i4} \bN^*_{j,n+4}\right) + (i\leftrightarrow j)\right],\nonumber\\
\label{nna-L-couplings}
\eea
and
\beq
O^{nna}_{Rijk} =-(O^{nna}_{Lijk})^*.
\label{nna-R-couplings}
\eeq

\subsection*{Neutralino-neutralino-$Z^0$}

The Lagrangian using four component spinor notation can be written as
\beq
\mathcal{L}^{nnz}= - \frac{g_2}{2} \ovl{{\widetilde \chi}^0_i} \gamma^{\mu} (O^{nnz}_{Lij} P_L + O^{nnz}_{Rij} P_R) {{\widetilde \chi}^0_j} Z^0_{\mu},
\label{Neutralino-neutralino-neutral-Z-boson}
\eeq

where
\bea
O^{nnz}_{Lij} = & & \eta_{i} \eta_{j} \frac{1}{2 \cos_{\theta_W}} \left(\bN_{i3} \bN^*_{j3} - \bN_{i4} \bN^*_{j4} + \bN_{i,m+7} \bN^*_{j,m+7}\right), \nonumber \\
O^{nnz}_{Rij} = & & \frac{1}{2 \cos_{\theta_W}} \left(-\bN^*_{i3} \bN_{j3} + \bN^*_{i4} \bN_{j4} - \bN^*_{i,m+7} \bN_{j,m+7}\right). 
\label{nnz-L-R-couplings}
\eea

\subsection*{Neutralino-chargino-charged scalar}
The Lagrangian using four component spinor notation can be written as
\beq
\mathcal{L}^{ncs}= - {\widetilde g} \ovl{{\widetilde \chi}_i} (O^{cns}_{Lijk} P_L + O^{cns}_{Rijk} P_R) {{\widetilde \chi}^0_j} S^+_k - {\widetilde g} \ovl{{\widetilde \chi}^0_i} (O^{ncs}_{Lijk} P_L + O^{ncs}_{Rijk} P_R) {{\widetilde \chi}_j} S^-_k,
\label{Neutralino-chargino-chargedscalar}
\eeq
where
\bea
{\widetilde g} O^{cns}_{Lijk} = & &\eta_{j} \left[\bRc_{k1} \left( -\frac{g_2}{\sqrt{2}} {\bU}^*_{i2} \bN^*_{j2} - \frac{g_1}{\sqrt{2}} \bU^*_{i2} \bN^*_{j1} + {g_2} \bU^*_{i1} \bN^*_{j3}\right) 
\right. \nonumber \\
& &+ \bRc_{k2} \left( \lm^m \bU^*_{i2} \bN^*_{j,m+4} - Y^{mn}_{\nu} \bU^*_{i,m+2} \bN^*_{j,n+4}\right) \nonumber \\
& &+ \bRc_{k,m+2} \left(Y^{mn}_e \bU^*_{i,n+2} \bN^*_{j3} - Y^{mn}_e \bU^*_{i2} \bN^*_{j,n+7} \right)\nonumber \\
& &+ \left.
\bRc_{k,m+5} \left( {g_2} {\bU}^*_{i1} \bN^*_{j,m+7} - \frac{g_2}{\sqrt{2}} \bU^*_{i,m+2} \bN^*_{j2} - \frac{g_1}{\sqrt{2}} \bU^*_{i,m+2} \bN^*_{j1}\right)\right],\nonumber \\
\nonumber\\
\label{cns-L-couplings}
{\widetilde g} O^{cns}_{Rijk} = & &\epsilon_{i} \left[\bRc_{k1} \left( \lm^m \bV_{i2} \bN_{j,m+4} -Y^{mn}_e \bV_{i,n+2} \bN_{j,m+7}\right) 
\right. \nonumber \\
& &+ \bRc_{k2} \left( \frac{g_2}{\sqrt{2}} {\bV}_{i2} \bN_{j2} + \frac{g_1}{\sqrt{2}} \bV_{i2} \bN_{j1} + {g_2} \bV_{i1} \bN_{j4}\right) \nonumber \\
& &+ \sqrt{2} {g_1} \bRc_{k,m+2} \bV_{i,m+2} \bN_{j1}\nonumber \\
& &+ \left.
\bRc_{k,m+5} \left(Y^{mn}_e \bV_{i,n+2} \bN_{j3} - Y^{mn}_{\nu} \bV_{i2} \bN_{j,n+4} \right)\right],\nonumber \\
\label{cns-R-couplings}
\eea
and
\beq
O^{ncs}_{Lijk} =(O^{cns}_{Rjik})^*, \quad O^{ncs}_{Rijk} =(O^{cns}_{Ljik})^*.
\label{ncs-L-R-couplings}
\eeq

\subsection*{Neutralino-chargino-$W$}
The Lagrangian using four component spinor notation can be written as
\beq
\mathcal{L}^{ncw}=  -  {g_2} \ovl{{\widetilde \chi}_i} \gamma^{\mu} (O^{cnw}_{Lij} P_L + O^{cnw}_{Rij} P_R) {{\widetilde \chi}^0_j} 
W^+_{\mu} -  {g_2} \ovl{{\widetilde \chi}^0_i} \gamma^{\mu} (O^{ncw}_{Lij} P_L + O^{ncw}_{Rij} P_R) {{\widetilde \chi}_j} W^-_{\mu}.
\label{Neutralino-chargino-W-boson}
\eeq
where
\bea{\label{ncw-L-R-couplings}}
O^{cnw}_{Lij} = & &- \epsilon_{i} \eta_{j} \left(\bV_{i1} \bN^*_{j2} - \frac{1}{\sqrt{2}}  \bV_{i2} \bN^*_{j4}\right), \nonumber \\
O^{cnw}_{Rij} = & &- \bU^*_{i1} \bN_{j2} - \frac{1}{\sqrt{2}}  \bU^*_{i2} \bN_{j3}- \frac{1}{\sqrt{2}} \bU^*_{i,n+2} \bN_{j,n+7}, 
\eea
and
\beq
O^{ncw}_{Lij} =(O^{cnw}_{Lji})^*, \quad O^{ncw}_{Rij} =(O^{cnw}_{Rji})^*.
\label{cnw-L-R-couplings}
\eeq
The factors $\eta_{j}$ and $\epsilon_{i}$ are the proper signs of neutralino and chargino masses \cite{Gunion-Haber-1}. They have values $\pm{1}$. 
\subsection*{Neutralino-quark-squark}
The Lagrangian using four component spinor notation can be written as
\beq
\mathcal{L}^{nq\q}=  -  {\widetilde g} \ovl{q_i} (O^{qn\q}_{Lijk} P_L + O^{qn\q}_{Rijk} P_R) {{\widetilde \chi}^0_j} \q_k -  {\widetilde g} \ovl{{\widetilde \chi}^0_i} (O^{nq\q}_{Lijk} P_L + O^{nq\q}_{Rijk} P_R) q_j \q^*_k.
\label{Neutralino-quark-squark}
\eeq
where
\beq
O^{qn\q}_{Lijk} =(O^{nq\q}_{Rjik})^*, \quad O^{qn\q}_{Rijk} =(O^{nq\q}_{Ljik})^*,
\label{q-n-squark-L-R-couplings}
\eeq
and
\bea
{\widetilde g} O^{nu\u}_{Lijk} = & & \Rsu_{km} \left(\frac{g_2}{\sqrt{2}} \bN^*_{i2} \Ru_{L_{jm}} + \frac{g_1}{3\sqrt{2}} \bN^*_{i1} \Ru_{L_{jm}}\right) + Y^{nm}_u \Rsu_{k,m+3} \bN^*_{i4} \Ru_{L_{jn}} , \nonumber \\
{\widetilde g} O^{nu\u}_{Rijk} = & & Y^{{mn}^*}_u \Rsu_{km} \bN_{i4} R^{u^*}_{R_{jn}} - \frac{4{g_1}}{3 \sqrt{2}} \Rsu_{k,m+3} \bN_{i1} R^{u^*}_{R_{jm}},\nonumber \\
{\widetilde g} O^{nd\d}_{Lijk} = & & \Rsd_{km} \left(-\frac{g_2}{\sqrt{2}} \bN^*_{i2} \Rd_{L_{jm}} + \frac{g_1}{3\sqrt{2}} \bN^*_{i1} \Rd_{L_{jm}}\right) + Y^{nm}_d \Rsd_{k,m+3} \bN^*_{i3} \Rd_{L_{jn}} , \nonumber \\
{\widetilde g} O^{nd\d}_{Rijk} = & & Y^{{mn}^*}_d \Rsd_{km} \bN_{i3} R^{d^*}_{R_{jn}} + \frac{2{g_1}}{3 \sqrt{2}} \Rsd_{k,m+3} \bN_{i1} R^{d^*}_{R_{jm}}.
\label{n-q-squark-L-R-couplings}
\eea
\section{The ${\widetilde \Sigma}^V_{ij}$ and ${\widetilde \Pi}^V_{ij}$ function}\label{self-energy-sigma-pi}
In this appendix we give the detail expressions for the renormalized self energy functions ${\widetilde \Sigma}^V_{ij}$ and ${\widetilde \Pi}^V_{ij}$.
The net result is
\bea\label{Sigma-Pi-part}
{\widetilde \Sigma}^V_{ij} = -\frac{1}{16 \pi^2}&&\left[ \frac{{\widetilde g}^2}{2} \sum_{r=1}^{8} \sum_{k=1}^{10}  \left(O^{nnh}_{Lkir} O^{nnh}_{Rjkr} + O^{nnh}_{Ljkr} O^{nnh}_{Rkir}\right) B_1(p^2,m^2_{\n_k},m^2_{S^0_r}) 
\right. \nonumber\\
- && \frac{{\widetilde g}^2}{2} \sum_{r=1}^{7} \sum_{k=1}^{10}  \left(O^{nna}_{Lkir} O^{nna}_{Rjkr} + O^{nna}_{Ljkr} O^{nna}_{Rkir}\right) B_1(p^2,m^2_{\n_k},m^2_{P^0_r})\nonumber\\
+ && {g^2_2} \sum_{k=1}^{10}  \left( O^{nnz}_{Lki} O^{nnz}_{Ljk} + O^{nnz}_{Rki} O^{nnz}_{Rjk} \right) {B_1(p^2,m^2_{\n_k},m^2_{Z^0_{\mu}})}\nonumber\\
+ && {2 g^2_2} \sum_{k=1}^{5}  \left( O^{cnw}_{Lki} O^{ncw}_{Ljk} + O^{cnw}_{Rki} O^{ncw}_{Rjk} \right) {B_1(p^2,m^2_{{\widetilde\chi}^{\mp}_k},m^2_{W^{\pm}_{\mu}})}\nonumber\\
+ && {{\widetilde g}^2} \sum_{r=1}^{7} \sum_{k=1}^{5} \left(O^{cns}_{Lkir} O^{ncs}_{Rjkr} + O^{ncs}_{Ljkr} O^{cns}_{Rkir}\right) {B_1(p^2,m^2_{{\widetilde\chi}^{\mp}_k},m^2_{S^{\pm}_r})}\nonumber\\
+ && 3 {{\widetilde g}^2} \sum_{r=1}^{6} \sum_{k=1}^{3} \left(O^{un\u}_{Lkir} O^{nu\u}_{Rjkr} + O^{nu\u}_{Ljkr} O^{un\u}_{Rkir}\right) {B_1(p^2,m^2_{u_k},m^2_{\u_r})}\nonumber\\
+ && \left.
3{{\widetilde g}^2} \sum_{r=1}^{6} \sum_{k=1}^{3} \left(O^{dn\d}_{Lkir} O^{nd\d}_{Rjkr} + O^{nd\d}_{Ljkr} O^{dn\d}_{Rkir}\right) {B_1(p^2,m^2_{d_k},m^2_{\d_r})} \right]
\,,
\eea
\bea\label{Pi-part}
\nonumber \\
{\widetilde \Pi}^V_{ij} = -\frac{1}{16 \pi^2}&&\left[ {{\widetilde g}^2} \sum_{r=1}^{8} \sum_{k=1}^{10} \frac{m_{\n_k}}{2} \left(O^{nnh}_{Lkir} O^{nnh}_{Ljkr} + O^{nnh}_{Rkir} O^{nnh}_{Rjkr}\right) B_0(p^2,m^2_{\n_k},m^2_{S^0_r}) 
\right. \nonumber\\
- &&{{\widetilde g}^2 } \sum_{r=1}^{7} \sum_{k=1}^{10} \frac{m_{\n_k}}{2} \left(O^{nna}_{Lkir} O^{nna}_{Ljkr} + O^{nna}_{Rkir} O^{nna}_{Rjkr}\right) B_0(p^2,m^2_{\n_k},m^2_{P^0_r})\nonumber\\
- && {2 g^2_2} \sum_{k=1}^{10} {m_{\n_k}} \left(O^{nnz}_{Lki} O^{nnz}_{Rjk} + O^{nnz}_{Ljk} O^{nnz}_{Rki}\right) {B_0(p^2,m^2_{\n_k},m^2_{Z^0_{\mu}})}\nonumber\\
- && {4 g^2_2} \sum_{k=1}^{5} {m_{\widetilde{\chi}^{\pm}_k}} \left(O^{cnw}_{Lki} O^{ncw}_{Rjk} + O^{cnw}_{Rki} O^{ncw}_{Ljk}\right) {B_0(p^2,m^2_{{\widetilde\chi}^{\mp}_k},m^2_{W^{\pm}_{\mu}})}\nonumber\\
+ && {{\widetilde g}^2} \sum_{r=1}^{7} \sum_{k=1}^{5} {m_{\widetilde{\chi}^{\pm}_k}} \left(O^{cns}_{Lkir} O^{ncs}_{Ljkr} + O^{ncs}_{Rjkr} O^{cns}_{Rkir}\right) {B_0(p^2,m^2_{{\widetilde\chi}^{\mp}_k},m^2_{S^{\pm}_r})}\nonumber\\
+ && 3{{\widetilde g}^2} \sum_{r=1}^{6} \sum_{k=1}^{3} {m_{u_k}} \left(O^{un\u}_{Lkir} O^{nu\u}_{Ljkr} + O^{un\u}_{Rkir} O^{nu\u}_{Rjkr}\right) {B_0(p^2,m^2_{u_k},m^2_{\u_r})}\nonumber\\
+ && \left.
3{{\widetilde g}^2} \sum_{r=1}^{6} \sum_{k=1}^{3} {m_{d_k}} \left(O^{dn\d}_{Lkir} O^{nd\d}_{Ljkr} + O^{dn\d}_{Rkir} O^{nd\d}_{Rjkr}\right) {B_0(p^2,m^2_{d_k},m^2_{\d_r})} \right]
\,.
\eea
Detail expressions for the couplings are given in appendix \ref{Feynman rules}. The $B_0,~B_1$ functions are given in appendix \ref{The $B_0$ and $B_1$ function}. The factor $3$ appearing in front of the quark-squark loop contributions signifies three variations of quark colour.
\section{The $B_0$ and $B_1$ function}\label{The $B_0$ and $B_1$ function}

The $B_0$ and $B_1$ functions are Passarino-Veltman \cite{Passarino-Veltman,Veltman-tHooft} functions defined in the notation of \cite{Hahn-Victoria} 

\bea{\label{Passarino-Veltman-Functions}}
\frac{i}{16 {\pi}^2} {B_0(p^2,m^2_{f^\prime_k},m^2_{b_r})} &=& {\mu^{4-D}}\int \frac{d^Dq}{(2\pi)^D} \frac{1}{(q^2-m^2_{f^\prime_k})((q+p)^2-m^2_{b_r})},\nonumber \\
\frac{i}{16 {\pi}^2} {B_{\mu}(p^2,m^2_{f^\prime_k},m^2_{b_r})} &=& {\mu^{4-D}}\int \frac{d^Dq}{(2\pi)^D} \frac{q_{\mu}}{(q^2-m^2_{f^\prime_k})((q+p)^2-m^2_{b_r})},\nonumber \\
{B_{\mu}(p^2,m^2_{f^\prime_k},m^2_{b_r})} &=& p_{\mu} {B_1(p^2,m^2_{f^\prime_k},m^2_{b_r})}.\nonumber \\
\eea

\section{Some Useful Relations}\label{Some Useful Relations}

\subsection*{Fermionic sector}

For neutralinos the following relations between mass and weak eigenstates are 
very useful

\bea\label{neutralino_mass-basis_weak-basis_relations}
& &{P_L}{\widetilde{B}}^0 = P_L N^*_{i1} \widetilde{\chi}^0_i, \quad\ {P_L}{\widetilde{W}}^0_3 = P_L N^*_{i2} \widetilde{\chi}^0_i, \quad\ {P_L}{\widetilde{H}}_j = P_L N^*_{i,j+2} \widetilde{\chi}^0_i,\nonumber \\
& &{P_L}{\nu}_k = P_L N^*_{i,k+7} \widetilde{\chi}^0_i, \quad\ {P_L}{\nu}^{c}_k = P_L N^*_{i,k+4} \widetilde{\chi}^0_i,\nonumber \\
& &{P_R}{\widetilde{B}}^0 = P_R N_{i1} \widetilde{\chi}^0_i, \quad\ {P_R}{\widetilde{W}}^0_3 = P_R N_{i2} \widetilde{\chi}^0_i, \quad\ {P_R}{\widetilde{H}}_j = P_R N_{i,j+2},\quad \nonumber \\
& &{P_R}{\nu}_k = P_R N_{i,k+7} \widetilde{\chi}^0_i, \quad\ {P_R}{\nu}^{c}_k = P_R N_{i,k+4} \widetilde{\chi}^0_i, \nonumber \\
& & \text{where} \quad j = 1,2 \quad \text{and} \quad k = 1,2,3,
\eea
with i varies from 1 to 10 and 
\beq\label{P-L-P-R}
P_{L}=\left(\frac{1-{\gamma^5}}{2}\right), \quad 
P_{R}=\left(\frac{1+{\gamma^5}}{2}\right).
\eeq

In terms of the four component spinors $\chi_i$ for charginos, the following 
relations between mass and weak eigenstates are very useful. 

\bea\label{chargino_mass-basis_weak-basis_relations}
& &{P_L}{\widetilde{W}} = P_L V^*_{i1} \widetilde{\chi}_i, \quad\ {P_L}{\widetilde{H}} = P_L V^*_{i2} \widetilde{\chi}_i, \quad\ {P_L}{l_k} = P_L U^*_{i,k+2} \widetilde{\chi}^c_i,\nonumber \\
& &{P_R}{\widetilde{W}} = P_R U_{i1} \widetilde{\chi}_i, \quad\ {P_R}{\widetilde{H}} = P_R U_{i2} \widetilde{\chi}_i, \quad\ {P_R}{l_k} = P_R V_{i,k+2} \widetilde{\chi}^c_i,\nonumber \\
& &{P_L}{\widetilde{W}^c} = P_L U^*_{i1} \widetilde{\chi}^c_i, \quad\ {P_L}{\widetilde{H}^c} = P_L U^*_{i2} \widetilde{\chi}^c_i, \quad\ {P_L}{l^c_k} = P_L V^*_{i,k+2} \widetilde{\chi}_i,\nonumber \\
& &{P_R}{\widetilde{W}^c} = P_R V_{i1} \widetilde{\chi}^c_i, \quad\ {P_R}{\widetilde{H}^c} = P_R V_{i2} \widetilde{\chi}^c_i, \quad\ {P_R}{l^c_k} = P_R U_{i,k+2} \widetilde{\chi}_i,\nonumber \\
\eea
where $k = 1,2,3$, and i varies from 1 to 5. The last six relations are for the charge-conjugated fields.

The four component neutralino, chargino and charge conjugated chargino spinors are respectively defined as
\bea
& &{\widetilde \chi}^0_i =
\left(\begin{array}{c}
\chi^0_i \\
\ovl{\chi^0_i}\\
\end{array}\right),\quad
{\widetilde \chi}_i =
\left(\begin{array}{c}
\chi^+_i \\
\ovl{\chi^-_i}\\
\end{array}\right),\quad
{\widetilde \chi}^c_i =
\left(\begin{array}{c}
\chi^-_i \\
\ovl{\chi^+_i}\\
\end{array}\right),\nonumber \\
\label{neutralino-chargino}
\eea
where $\chi^0_i$ and $\chi^{\pm}_i$ are two component neutral and charged spinors, respectively.

\subsection*{Scalar sector}

The relations between weak and mass eigenstates for neutral scalar, neutral pseudoscalar and charged scalar 
are given by eqs.(\ref{pseudoscalar-mass-basis}), (\ref{scalar-mass-basis}), and (\ref{charged-scalar-mass-basis}). 


\end{document}